\documentclass[a4paper,11pt] {article}
\pdfoutput=1 % if your are submitting a pdflatex (i.e. if you have
\usepackage{jheppub} % for details on the use of the package, please
\usepackage{amsthm,enumerate,bbold}
\usepackage[utf8]{inputenc}
\usepackage{lmodern}
\allowdisplaybreaks
\newcommand{\al}{\alpha}
\newcommand{\be}{\beta}
\newcommand{\de}{\delta}

\newcommand{\vep}{\varepsilon}
\newcommand{\ga}{\gamma}

\newcommand{\la}{\lambda}

\newcommand{\si}{\sigma}
\renewcommand{\th}{\theta}

\newcommand{\Ga}{\Gamma}

\let\wt\widetilde

\newcommand{\tb}{\widetilde{b}}
\newcommand{\tc}{\widetilde{c}}

\newcommand{\CC}{{\mathbb C}}

\newcommand{\RR}{{\mathbb R}}

\newcommand{\cF}{{\mathcal F}}
\newcommand{\cG}{{\mathcal G}}
\newcommand{\cH}{{\mathcal H}}
\newcommand{\cI}{{\mathcal I}}

\newcommand{\cK}{{\mathcal K}}
\newcommand{\cL}{{\mathcal L}}

\newcommand{\cP}{{\mathcal P}}

\newcommand{\cT}{{\mathcal T}}

\newcommand{\pd}{\partial}

\newcommand{\id}{\mathbb{1}}
\newcommand{\noi}{\noindent}

\let\ds\displaystyle

\newcommand{\mss}{\kern 1pt}

\renewcommand{\le}{\leqslant}
\renewcommand{\ge}{\geqslant}
\newcommand{\tends}[1]{\bbuildrel{\hbox to 2em{\rightarrowfill}}_{#1}^{}}

\newcommand{\sgn}{\operatorname{sgn}}

\newcommand{\tr}{\operatorname{tr}}

\newcommand{\diag}{\operatorname{diag}}

\renewcommand{\Re}{\operatorname{Re}}
\newcommand{\iu}{\mathrm i}
\newcommand{\e}{\mathrm e}
\newcommand{\diff}{\mathrm{d}}

\newcommand{\su}{\mathrm{su}}

\newcommand{\SU}{\mathrm{SU}}
\renewcommand{\Im}{\operatorname{Im}}

\newcommand{\en}{\enspace}

\newcommand{\Int}[1]{\,\mathop{\!#1}\limits^{\lower1ex\hbox{$\scriptstyle\circ$}}{}}

\theoremstyle{remark}
\newtheorem{remark}{Remark}

\def\clap#1{\hbox to 0pt{\hss#1\hss}}

\title{Integrable coupled massive Thirring model with field values in a Grassmann algebra}
\author[a]{B. Basu-Mallick,}
\author[a]{F. Finkel,}
\author[a,1]{A. Gonz\'alez-L\'opez,\note{Corresponding author.}}
\author[b]{D. Sinha}
\affiliation[a]{Departamento de F\'\i sica Te\'orica, Universidad Complutense de Madrid, Plaza de
  las Ciencias 1,\\28040 Madrid, SPAIN}
\affiliation[b]{Theory Division, Saha Institute of Nuclear Physics, HBNI, 1/AF Bidhannagar,\\
  Kolkata 700064, INDIA}

\emailAdd{bireswar.basumallick@saha.ac.in} \emailAdd{ffinkel@ucm.es} \emailAdd{artemio@ucm.es}
\emailAdd{debdeep.sinha@saha.ac.in}

\abstract{%
  A coupled massive Thirring model of two interacting Dirac spinors in $1+1$ dimensions with
  fields taking values in a Grassmann algebra is introduced, which is closely related to a
  $\SU(1,1)$ version of the Grassmannian Thirring model also introduced in this work. The Lax pair
  for the system is constructed, and its equations of motion are obtained from a zero curvature
  condition. It is shown that the system possesses several infinite hierarchies of conserved
  quantities, which strongly confirms its integrability. The model admits a canonical formulation
  and is invariant under space-time translations, Lorentz boosts and global $\mathrm U(1)$ gauge
  transformations, as well as discrete symmetries like parity and time reversal. The conserved
  quantities associated to the continuous symmetries are derived using Noether's theorem, and
  their relation to the lower-order integrals of motion is spelled out. New nonlocal integrable
  models are constructed through consistent nonlocal reductions between the field components of
  the general model. The Lagrangian, the Hamiltonian, the Lax pair and several infinite
  hierarchies of conserved quantities for each of these nonlocal models are obtained substituting
  its reduction in the expressions of the analogous quantities for the general model. It is shown
  that, although the Lorentz symmetry of the general model breaks down for its nonlocal
  reductions, these reductions remain invariant under parity, time reversal, global $\mathrm U(1)$
  gauge transformations and space-time translations.}

\begin{document}

\keywords{Integrable Field Theories [100], Integrable Hierarchies [25]}

\maketitle
\flushbottom

\section{Introduction}

The massive Thirring model (MTM) has attracted considerable attention in the literature as a rare
example of an exactly solvable relativistic model in $1+1$ dimensions, which is relevant in
different branches of physics. The MTM was originally introduced as a toy model in the realm of
quantum field theory, with the purpose of understanding the non-perturbative physical phenomena
arising in realistic $(3+1)$-dimensional systems~\cite{Th58}. Further developments such as the
Coleman correspondence~\cite{Co75} between the quantum sine-Gordon model and the zero charge
sector of the quantum MTM stimulated the study of the Thirring model as an integrable quantum
field theory. In fact, subsequent work established the complete integrability of the MTM both in
its quantum~\cite{Th81,KS82,Bh05} and classical variants~\cite{Mi76,Or76,IK83,IK78}. More
precisely, at the classical level there exist two versions of the MTM, depending on whether the
field variables take values in the complex numbers field or in a Grassmann algebra. The first
version, often referred to as the bosonic massive Thirring model (BMTM), has important physical
applications including pulse propagation in Bragg nonlinear optical
media~\cite{WC82,CJ89,AW89,ESSKS96,ESS99} and dipole-dipole interactions among many-body bosonic
atoms~\cite{LMW93}. A great deal of research has been devoted to the BMTM applying well-known
techniques such as the inverse scattering transform, Bäcklund or Darboux transformations, and
various types of soliton solutions and rogue waves thereof have been
obtained~\cite{Mi76,Or76,KM77,KMI79,KN77,Wa83,PS19,Pr81,De15,DWA15,GWCH17,YBPCMB21}. Further
interesting results include the proof of the existence of soliton solutions in a nonvanishing
background~\cite{BG87,BGK93,BG93} and the construction of general bright and dark soliton
solutions via KP hierarchy reductions~\cite{CF21}, or the study of the integrability of the model
in the presence of defects~\cite{AGYZ11,Ag12} and of balanced loss and gain~\cite{ABS19}.

The study of the integrability of the classical MTM with fields taking values in a Grassmann
algebra (GMTM for short) is important in the context of quantization, in which case the fermionic
fields are taken as operators satisfying canonical anticommutation relations. The integrability of
the GMTM at the classical level was established early on in Refs.~\cite{IK83,IK78}. The Lax pair
for this model was constructed in the latter reference, and an infinite number of conserved
quantities were obtained. Furthermore, by employing the inverse scattering transform it was shown
that the usual soliton solution is absent in this model. A study of the GMTM in the presence of
defects was also undertaken in Refs.~\cite{AGYZ11,Ag12}.

A coupled BMTM describing the interaction between two independent self-interacting Dirac spinors
was recently introduced in Ref.~\cite{BS23}, where the model's Lagrangian and Hamiltonian were
constructed. It was shown that the action of the system is invariant under parity, time reversal,
global $\mathrm U(1)$ gauge transformations and Lorentz boosts, as is the case for the original
MTM. The new model possesses a Lax pair which yields the equations of motion as a zero curvature
condition. The linear equations associated with this Lax pair can be used to construct an infinite
number of conserved quantities, which confirms the integrability of the system. A detailed study
of the complete integrability (in the Liouville sense) of the coupled BMTM was also carried out in
Ref.~\cite{BS23}, and some novel nonlocal nonlinear integrable systems related to different
reductions of the coupled BMTM were constructed and analyzed.

The integrability of both the bosonic and Grassmannian versions of the original MTM, as well as
the new coupled MTM introduced in Ref.~\cite{BS23}, strongly suggest that this might also be the
case for the Grassmannian version of the latter model. In fact, the main aim of the present paper
is to investigate the integrability properties of the classical version of the coupled MTM with
fields taking values in a Grassmann algebra. To this end, we first construct the model's Lax pair,
which consists of $3\times 3$ matrices instead of $2\times 2$ matrices as is the case in its
bosonic version. We then show that the compatibility condition between the two linear Lax
equations yields the equations of motion for the coupled GMTM via a zero curvature condition.

Another interesting outcome of the present investigation is that the original Lagrangian for this
coupled GMTM can be written in a new form by introducing a pair of new Grassmann fields related to
the old ones through a rotation by an angle $-\pi/4$ in the internal space of the field variables.
We show that in this form the model is closely related to a $\SU(1,1)$ generalization of the GMTM
model that, to the best of our knowledge, has not been previously discussed in the literature. On
the other hand, the quantized version of this new form of the Lagrangian of the coupled GMTM may
also be interpreted as describing the mutual interaction between a self-interacting Thirring fermion
and a ghost Thirring fermion of the same mass but with opposite sign of the coupling constant. It
should be noted in this respect that this type of ghost fields has recently been considered in the
literature in the context of the bosonic field theory of quantized gravity~\cite{GSTZ21}.

Apart from the Lax pair, a fundamental aspect for the integrability of a given system is the
existence of conserved quantities. The construction of the conserved quantities of the new model
is accomplished by exploiting the linear equations coming from the Lax pair and the zero curvature
condition. In this way several infinite hierarchies of conserved quantities are obtained, which
ensure the integrability of the system. As in the case of the ordinary GMTM, some of the conserved
quantities of the new coupled GMTM are local whereas others turn out to be nonlocal (by contrast,
all the conserved quantities of the coupled BMTM are local). The Hamiltonian formulation of the
coupled GMTM is presented following the standard canonical formalism for Grassmann-valued field
theories~\cite{Ca76,Ma85}. The action for the coupled GMTM is shown to be manifestly invariant
under space-time translations and $\mathrm U(1)$ global gauge transformations, and the
corresponding conserved charges are obtained through Noether’s theorem. These conserved charges,
which may readily be identified as the system's Hamiltonian, total momentum and charge, are in
fact related to the local integrals of motion constructed using the Lax pair formalism. Finally,
the Noether current arising from the invariance of the system under Lorentz boosts is also
obtained. Apart from these continuous symmetries, the system is also invariant under several
discrete symmetries, namely parity (${\cal P}$) and time reversal (${\cal T}$), and hence under
their composition (${\cal PT}$). Interestingly, the lower-order integrals of motion of the coupled
GMTM are found to Poisson commute among themselves, a fact strongly suggesting that the system
might be completely integrable in Liouville's sense as in the case of the coupled BMTM. However, a
full proof of the complete integrability for the coupled GMTM is not considered in the present
study and will be the subject of future work.

A systematic study of the reductions of the field components of the coupled GMTM shows that its
Lax pair can be used to generate new integrable systems. Apart from the ordinary GMTM, we obtain
two nonlocal reductions with real reverse space and real reverse time, producing new integrable
systems. These two models are nonlocal in the sense that the value of the interaction potential at
$(x,t)$ depends on the value of the fields at $(-x,t)$ (space inversion) or at $(x,-t)$ (time
inversion). It should be mentioned here that the type of nonlocal integrable models discussed in
the present paper were first introduced in the context of the nonlinear Schrödinger equation
(NLSE) with a space inversion interaction~\cite{AM13}. Since then, a vast amount of research has
been carried out on several types of nonlocal integrable dynamical systems, including different
generalized versions of the NLSE ~\cite{SMMC14, SG17, AM17, KS15, SG15,GS17,Si22}, nonlocal
derivative NLSE~\cite{SSZ19,Zh18} and nonlocal sine-Gordon model~\cite{XZS22}. To the best of our
knowledge, however, such nonlocal integrable models had not been constructed so far for field
theories with Grassmann-valued variables. In our present study, the Lax pair and integrals of
motion for the new nonlocal integrable models are simply constructed by reduction of the analogous
expressions for the coupled GMTM. The Lagrangian and the corresponding Hamiltonian are also
derived in the same way. It is also shown that the new nonlocal reductions break the Lorentz
invariance of the coupled GMTM. However, they remain invariant under global $\mathrm U(1)$ gauge
transformations and discrete space-time symmetries (parity, time reversal and their composition).

The paper's organization is as follows. In the next section we introduce the coupled MTM with
field values in a Grassmann algebra. The Lax pair of the model is constructed, and the equations
of motion are derived as the corresponding zero curvature condition. Sections~\ref{sec.NCQ}
and~\ref{sec.LCQ} respectively deal with the construction of several infinite hierarchies of
nonlocal and local conserved quantities by using the Lax equations and the zero curvature
condition. In Section~\ref{sec.CF} the canonical formulation of the model is presented. The
continuous symmetries of the system ---space-time translations, global $\mathrm U(1)$ gauge
transformations and Lorentz boosts--- are discussed, and the corresponding conserved quantities
are obtained by applying Noether's theorem. The invariance of the system under discrete symmetries
---parity and time reversal--- is also discussed in this section. The nonlocal integrable models
obtained by considering different reductions between the field components of the coupled GMTM are
presented in Section~\ref{sec.NRS}. Finally, in the last section we make a brief summary of the
results obtained and outline some future developments.

\section{The model}\label{sec.Lax}

\subsection{Definition and Lax pair}

A (bosonic) coupled MTM in $1+1$ dimensions incorporating the interaction between two Lorentz
spinors $\psi$ and $\phi$ was recently introduced in Ref.~\cite{BS23}. The Lagrangian density for
this model is given by
\begin{align}
  \cL&=\frac{1}{2}\bar{\phi}\big(\iu\gamma^{\mu}\pd_{\mu}-m\big)\psi
       +\frac{1}{2}\bar{\psi}\big(\iu\gamma^{\mu}\pd_{\mu}-m\big)\phi
       -\frac{1}{2}\bar{\phi}\big(\iu\gamma^{\mu}\overset{\leftarrow}\pd_{\mu}+m\big)\psi
       -\frac{1}{2}\bar{\psi}\big(\iu\gamma^{\mu}\overset{\leftarrow}\pd_{\mu}+m\big)\phi
       \nonumber\\
     &\quad+\frac{g}{2}\big[(\bar{\phi}\gamma_{\mu}\psi)(\bar{\phi}\gamma^{\mu}\psi)
       +(\bar{\psi}\gamma_{\mu}\phi)(\bar{\psi}\gamma^{\mu}\phi)\big],
       \label{L}
\end{align} 
where $\phi=(\phi_1, \phi_2)^{\mathsf T}$, $\bar{\phi}=\phi^{\dagger}\gamma^0$ (and similarly for
$\psi$), the two-dimensional gamma matrices are taken as $\ga^0=\si_x$, $\ga^1=-\iu\si_y$, and the
spacetime metric tensor is $\diag(1,-1)$. The last term in square brackets represents the
interaction between the fields $\psi$ and $\phi$, $g$ being the (real) coupling constant. In the
present study we consider the fields $\phi_{i}, \psi_{j}$ (with $i, j=1,2$) as functions taking
values in the odd (fermionic) sector of an (infinite-dimensional) Grassmann algebra $\mathbf{G}$.
In other words, the fields $\phi_i$ and $\psi_j$ anticommute among themselves and with their
conjugates $\phi_i^*$ and $\psi_j^*$, where $``*"$ denotes the involution (complex conjugation) in
$\mathbf{G}$~\cite{De92}. The Euler--Lagrange equations of motion for the fields $\phi$ and $\psi$
are thus
\begin{align}
  &\big(\iu\gamma^{\mu}\pd_{\mu}-m\big)\phi+g (\bar{\psi}\gamma_{\mu}\phi)\gamma^{\mu}\phi=0,
    \label{eqphi}\\
  &\big(\iu\gamma^{\mu}\pd_{\mu}-m\big)\psi+g (\bar{\phi}\gamma_{\mu}\psi)\gamma^{\mu}\psi=0,
    \label{eqpsi}
\end{align}
while the fields $\phi^*$ and $\psi^*$ obey the conjugate equations
\begin{align}
  &{\bar\phi}\big(\iu\gamma^{\mu}\overset {\leftarrow}\pd_{\mu}+m\big)-g{\bar\phi}
    \gamma^{\mu}(\bar{\phi}\gamma_{\mu}\psi)=0,
    \label{eqphic}\\
  &\bar{\psi}\big(\iu\gamma^{\mu}\overset {\leftarrow}\pd_{\mu}+m\big)-g{\bar\psi}
    \gamma^{\mu}(\bar{\psi}\gamma_{\mu}\phi)=0.
    \label{eqpsic}
\end{align}
These equations are clearly invariant under the exchange $\phi\leftrightarrow\psi$. Moreover, for
$g=0$ the fields $\phi$ and $\psi$ both satisfy the free Dirac equation, while for $\psi=\phi$
Eqs.~\eqref{eqphi} and \eqref{eqpsi} reduce to the equation of motion of the original
(Grassmannian) MTM.

Before proceeding further it is convenient to perform the change of scale
\begin{equation}
  x\to \frac{1}{m}x,\quad t\to \frac{1}{m}t,\quad \phi_j
  \to \sqrt{\frac{m}{2|g|}}\phi_j,\quad \psi_j\to \sqrt{\frac{m}{2|g|}}\psi_j,
  \qquad j=1,2,
  \label{cv}
\end{equation}
which transforms Eqs.~\eqref{eqphi}-\eqref{eqpsi} into
\begin{equation}
  \begin{aligned}
    &\iu\big(\pd_t+\pd_x\big)\phi_1-\phi_2-2\vep (\psi_2^{*}\phi_2)\phi_1=0,\\
    &\iu\big(\pd_t-\pd_x\big)\phi_2-\phi_1-2\vep (\psi_1^{*}\phi_1)\phi_2=0\\
    &\iu\big(\pd_t+\pd_x\big)\psi_1-\psi_2-2\vep (\phi_2^{*}\psi_2)\psi_1=0,\\
    &\iu\big(\pd_t-\pd_x\big)\psi_2-\psi_1-2\vep (\phi_1^{*}\psi_1)\psi_2=0
  \end{aligned}
  \label{eqsm}
\end{equation}
with $\vep:=-\sgn g$. The component form of the conjugate equations \eqref{eqphic}-\eqref{eqpsic}
is easily obtained by taking the complex conjugate of Eqs.~\eqref{eqsm}. Moreover, since the
change of variables $\phi\to-\phi$ (or $\psi\to-\psi$) reverses the sign of the term proportional
to $\vep$ in Eqs.~\eqref{eqsm}, we shall take from now on $\vep=1$ without loss of generality.

In order to study the integrability properties of the coupled GMTM~\eqref{eqsm} we shall employ
the zero curvature formulation, which relies on constructing a Lax pair for the system. The $U$,
$V$ matrices in this Lax pair can be expressed as $U=U_+$ and $V=U_-$, with
\begin{equation}
  U_{\pm}=
  \begin{pmatrix}
    \iu\,\rho_{\mp}+\frac{\iu}{2}(\la^2\mp\la^{-2})& 0 &-r_1^{\mp}\\[3pt]
    0 & -\iu\,\rho_{\mp} +\frac{\iu}{2}(\la^2\mp\la^{-2}) & -r_2^{\pm}\\[3pt]
    r_2^{\mp}& r_1^{\pm}&\iu(\la^2\mp\la^{-2})
  \end{pmatrix},
  \label{u}
\end{equation}
where
\begin{equation}\label{rhorr}
  \rho_{\pm}=\phi^{*}_2\psi_2\pm\phi^{*}_1\psi_1,\qquad r_1^{\pm}=-\iu\big(\la
  \phi^*_2\pm\la^{-1}\phi^*_1\big),\qquad r_2^{\pm}=\iu\big(\la \psi_2\pm\la^{-1}\psi_1\big)
\end{equation}
and $\la\in\CC$ is a spectral parameter independent of $x$ and $t$. The linear equations of the
Lax pair associated to the latter matrices are
\begin{equation}
  w_x=Uw,\qquad w_t=Vw,
  \label{uvn}
\end{equation}
where $w(x,t)$ is the three-component column vector $w=(w_1, w_2, w_3)^{\mathsf T}$,
$w_x=\frac{\pd w}{\pd x}$, and $w_t=\frac{\pd w}{\pd t}$. Note that from the form of the matrices
$U$ and $V$ it follows that the auxiliary field components $(w_1,w_2)$ and $w_3$ must have
opposite parity. The compatibility condition $w_{xt}=w_{tx}$ of the equations in~\eqref{uvn}
yields the zero curvature condition
\begin{equation}
  U_t-V_x+[U,V]=0,
  \label{zc}
\end{equation}
where $[U,V]$ is the usual commutator. It can be readily checked that, when written in component
form, Eq.~\eqref{zc} reduces to equations~\eqref{eqsm}. This shows that the equations of motion of
the coupled GMTM are indeed obtained as the zero curvature condition~\eqref{zc} of the Lax
pair~\eqref{uvn}.

\subsection{Connection to the $\SU(1,1)$ Thirring model}

The non-standard form of the kinetic energy terms in the Lagrangian~\eqref{L} can be easily
remedied by performing a rotation of angle $-\pi/4$ in internal (field) space, i.e., introducing a
pair of new two-component fermionic fields $(\Phi,\Psi)$ through the relation
\begin{equation}\label{PhiPsi}
  \phi=\frac1{\sqrt 2}\,(\Phi-\Psi),\qquad \psi=\frac1{\sqrt 2}\,(\Phi+\Psi).
\end{equation}
Indeed, in terms of the new fields $(\Phi,\Psi)$ the Lagrangian~\eqref{L} becomes
\begin{equation}\label{Ldiag}
  \cL=\cL_{\text{Th}}(\Phi;m,g/2)-\cL_{\text{Th}}(\Psi;m,-g/2)+\cL_{\text{I}},
\end{equation}
where
\begin{equation}\label{Lalt}
  \cL_{\text{Th}}(\chi;m,g)=\frac{1}{2}\bar{\chi}\big(\iu\gamma^{\mu}\pd_{\mu}-m\big)\chi
  -\frac{1}{2}\bar{\chi}\big(\iu\gamma^{\mu}\overset{\leftarrow}\pd_{\mu}+m\big)\chi
  +\frac{g}{2}(\bar{\chi}\gamma_{\mu}\chi)(\bar{\chi}\gamma^{\mu}\chi)
\end{equation}
is the usual (Grassmannian) Thirring Lagrangian, and the interaction Lagrangian~$\cL_\text{I}$ is
given by
\begin{equation}\label{LI}
  \cL_{\text{I}}=\frac{g}4\Big[(\bar{\Phi}\ga_\mu\Psi)(\bar{\Phi}\ga^\mu\Psi)+
  (\bar{\Psi}\ga_\mu\Phi)(\bar{\Psi}\ga^\mu\Phi)-2(\bar{\Phi}\ga_\mu\Phi)(\bar{\Psi}\ga^\mu\Psi)
  -2(\bar{\Phi}\ga_\mu\Psi)(\bar{\Psi}\ga^\mu\Phi) \Big].
\end{equation}

The minus sign in front of the second Thirring Lagrangian in Eq.~\eqref{Lalt} suggests a
connection of the present model with an $\SU(1,1)$ version of the massive Thirring model that we
shall now make explicit. To this end, we regard the pair $F=(\Phi,\Psi)^{\mathsf T}$ as a
(Grassmann-valued) vector field whose components $F^1=\Phi$ and $F^2=\Psi$ transform under the
fundamental representation of the $\SU(1,1)$ group. This is of course motivated by the fact that
the kinetic energy and mass terms of the Lagrangian~\eqref{Ldiag}-\eqref{LI} can be written as
\begin{equation}\label{KSU11}
  \frac12\eta_{ab}\Big[\bar F^{a}(\iu\ga_\mu\pd_\mu-m)F^b -\bar
  F^a\big(\iu\gamma^{\mu}\overset{\leftarrow}\pd_{\mu}+m\big)F^b \Big]\equiv \cK_{\SU(1,1)},
\end{equation}
where (as in what follows) we are implicitly summing over repeated indices,
$\bar F^ a=(F^a)^\dagger\ga^0$ and $\eta_{11}=-\eta_{22}=1$, $\eta_{12}=\eta_{21}=0$. This term is
easily seen to be invariant under the global $\SU(1,1)$ transformation $F\mapsto AF$, where $A$ is
a $2\times2$ complex matrix satisfying $A^\dagger\eta A=\eta$ and $\det A=1$. More generally,
inspired by the definition of the $\SU(N)$ (massless) Thirring model in Ref.~\cite{Zi02}, we
define the $\SU(1,1)$ massive Thirring model Lagrangian as
\begin{equation}
  \label{LSU11}
  \cL_{\SU(1,1)}= \cK_{\SU(1,1)}+g_0 J^0_\mu J^{0\mu}
  +g_1\eta_{bc}\eta_{ad}(\bar F^a\ga_\mu F^b)(\bar F^c\ga^\mu F^d),
\end{equation}
where
\[
  J^0_\mu=\eta_{ab}\bar F^a\ga_\mu F^b,\qquad \mu=0,1,
\]
is the $\mathrm U(1,1)$ current and $g_0,g_1$ are arbitrary real parameters. It is straightforward
to check that $\cL_{\SU(1,1)}$ is real, due to the identities $(\ga^0)^*=\ga^0$ and
$\ga^0\ga^*_\mu\ga^0=\ga_\mu$ satisfied by the gamma matrices. The Lagrangian~\eqref{LSU11} is
also readily seen to be invariant under the global $\SU(1,1)$ transformation
\[
  F\mapsto AF,\qquad A\in\SU(1,1),
\]
since both interaction terms in Eq.~\eqref{LSU11} are in fact separately invariant. The
Lagrangian~\eqref{LSU11} can also be written in terms of the three currents
\begin{equation}\label{Ji}
  J^i_\mu=\bar F^{a}(\eta t^i)_{ab}\ga_\mu F^b,\qquad i=1,2,3,
\end{equation}
associated to the $\su(1,1)$ generators\footnote{Here, and in what follows, we shall use the term
  ``generator'' adhering to the physicists' convention, according to which the one-parameter group
  generated by a generator $X$ is $\e^{\iu t X}$ instead of $\e^{t X}$ (with $t\in\RR$ the group
  parameter).} $t^i$ ($i=1,2,3$) in the fundamental representation, which we shall take as
\[
  t^1=\iu\si^x,\qquad t^2=\iu\si^y,\qquad t^3=\si^z.
\]
Note that $t^1$ and $t^2$ are anti-Hermitian while $t^3$ is Hermitian, and that
\[
  \tr t^i=0,\quad (t^i)^\dagger\eta-\eta t^i=0,\qquad i=1,2,3.
\]
Indeed, the trace formulas
\[
  \tr(t^it^j)=2\vep_i\de_{ij},\qquad \text{with}\quad \vep_3=-\vep_1=-\vep_2=1,
\]
imply the identity
\[
  \vep_i\eta_{a'a}\eta_{c'c}t^i_{ab}t^i_{cd}=2\eta_{a'd}\eta_{bc'}-\eta_{a'b}\eta_{c'd},
\]
and hence
\begin{equation}\label{su11trs}
  \vep_iJ^i_\mu J^{i\mu}=2\eta_{a'd}\eta_{bc'}(\bar F^{a'}\ga_\mu F^b)
  (\bar F^{c'}\ga^\mu F^d)-J^0_\mu J^{0\mu}.
\end{equation}
From this equation it immediately follows that the $\SU(1,1)$ Thirring model
Lagrangian~\eqref{LSU11} can be written as
\begin{equation}\label{LSU11c}
  \cL_{\SU(1,1)}=\cK_{\SU(1,1)}+\bigg(g_0+\frac{g_1}2\bigg)J^0_\mu J^{0\mu} +g_1\vep_i J^i_\mu
  J^{i\mu}.
\end{equation}
Note that, as is the case for the $\SU(N)$ Thirring model studied in Ref.~\cite{Zi02}, the last
term has the same structure as the $\su(1,1)$ Casimir $\vep_i t^it^i$.
  
In order to elucidate the precise relation between the $\SU(1,1)$ Thirring model~\eqref{LSU11c}
(or \eqref{LSU11}) and our model (cf.~Eqs.~\eqref{Ldiag}-\eqref{LI}), it suffices to expand the
interaction terms in Eq.~\eqref{LSU11}, namely:
\begin{align}
  J_\mu^0J^{0\mu}
  &=(\bar\Phi\ga_\mu\Phi-\bar\Psi\ga_\mu\Psi)(\bar\Phi\ga^\mu\Phi-\bar\Psi\ga^\mu\Psi)\notag\\
  &=(\bar\Phi\ga_\mu\Phi)(\bar\Phi\ga^\mu\Phi)+(\bar\Psi\ga_\mu\Psi)(\bar\Psi\ga^\mu\Psi)
    -2(\bar\Phi\ga_\mu\Phi)(\bar\Psi\ga^\mu\Psi),\label{J0}\\
  \eta_{bc}\eta_{ad}(\bar F^a\ga_\mu F^b)(\bar F^c\ga^\mu F^d)
  % &=
  % \eta_{bb}\eta_{aa}(\bar F^a\ga_\mu F^b)(\bar F^b\ga^\mu F^a)\\
    &=
      (\bar\Phi\ga_\mu\Phi)(\bar\Phi\ga^\mu\Phi)+(\bar\Psi\ga_\mu\Psi)(\bar\Psi\ga^\mu\Psi)
      -2(\bar\Phi\ga_\mu\Psi)(\bar\Psi\ga^\mu\Phi).\label{trFF}
\end{align}
We thus find that the $\SU(1,1)$ Thirring model Lagrangian~\eqref{LSU11} ---or, equivalently,
\eqref{LSU11c}--- can be written as
\begin{multline}\label{LSU11f}
  \cL_{\SU(1,1)}=\cK_{\SU(1,1)}
  +(g_0+g_1)\big[(\bar\Phi\ga_\mu\Phi)(\bar\Phi\ga^\mu\Phi)
  +(\bar\Psi\ga_\mu\Psi)(\bar\Psi\ga^\mu\Psi)
  \big]\\-2g_0(\bar\Phi\ga_\mu\Phi)(\bar\Psi\ga^\mu\Psi)
  -2g_1(\bar\Phi\ga_\mu\Psi)(\bar\Psi\ga^\mu\Phi).
\end{multline}
It is apparent that the terms
\[
  (\bar\Phi\ga_\mu\Psi)(\bar\Phi\ga^\mu\Psi)+(\bar\Psi\ga_\mu\Phi)(\bar\Psi\ga^\mu\Phi)
\]
in the Lagrangian~\eqref{Ldiag}-\eqref{LI} of the coupled GMTM do not appear in
Eq.~\eqref{LSU11f}, nor is it possible to reproduce the coefficients of the remaining terms
in~\eqref{LI} for any choice of $g_0$ and $g_1$. We thus conclude that the
Lagrangian~\eqref{Ldiag}-\eqref{LI} is \emph{not} of the $\SU(1,1)$ invariant form~\eqref{LSU11}.
On the other hand, \eqref{Ldiag}-\eqref{LI} can be expressed in terms of $\su(1,1)$ currents in a
remarkably simple way. Indeed, Eq.~\eqref{J0} and the identity
\begin{align*}
  % J^0_\mu J^{0\mu}
  % &=
  %   (\bar\Phi\ga_\mu\Phi-\bar\Psi\ga_\mu\Psi)(\bar\Phi\ga^\mu\Phi-\bar\Psi\ga^\mu\Psi)\\
  % &=(\bar\Phi\ga_\mu\Phi)(\bar\Phi\ga^\mu\Phi)+(\bar\Psi\ga_\mu\Psi)(\bar\Psi\ga^\mu\Psi)
  %   -2(\bar\Phi\ga_\mu\Phi)(\bar\Psi\ga_\mu\Psi),\\
  J^1_\mu J^{1\mu}
  &=
    -(\bar\Phi\ga_\mu\Psi-\bar\Psi\ga_\mu\Phi)(\bar\Phi\ga^\mu\Psi-\bar\Psi\ga^\mu\Phi)\\
  &=-(\bar\Phi\ga_\mu\Psi)(\bar\Phi\ga^\mu\Psi)-(\bar\Psi\ga_\mu\Phi)(\bar\Psi\ga^\mu\Phi)
    +2(\bar\Phi\ga_\mu\Psi)(\bar\Psi\ga_\mu\Phi),
\end{align*}
imply that
\begin{equation}\label{LJ1}
  \cL=\cK_{\SU(1,1)}+\frac g4\Big(J^0_\mu J^{0\mu}-J^1_\mu J^{1\mu}\Big).
\end{equation}
It is straightforward to check that each $\su(1,1)$ current $J^k_\mu$ is invariant under the
one-parameter group $\e^{\iu\th t^k}$ (with $\th\in\RR$) generated by its corresponding generator
$t^k\in\su(1,1)$. Indeed, under a general $\SU(1,1)$ transformation $F\mapsto AF$ the current
$J^k_\mu$ is mapped to
\[
  \bar F^{a}(A^\dagger\eta t^kA)_{ab}\ga_\mu F^b,
\]
and we have
\[
  \e^{-\iu\th (t^k)^\dagger}\eta t^k\e^{\iu\th t^k}=\eta t^k,\qquad k=1,2,3.
\]
For instance, for the current $J^1_\mu$ 
\begin{align*}
  \e^{-\iu\th (t^1)^\dagger}\eta t^1\e^{\iu\th t^1}
  &=\e^{-\th\si^x}(\iu\si^z\si^x)\e^{-\th\si^x}
    =-\e^{-\th\si^x}\si^y\e^{-\th\si^x}\\
  &=-(\cosh\th-\sinh\th\,\si^x)\si^y(\cosh\th-\sinh\th\,\si^x)=-\si^y
    =\iu\si^z\si^x=\eta t^1.
\end{align*}
Since the first two terms in the Lagrangian~\eqref{LJ1} are obviously $\SU(1,1)$ invariant, it
follows from the previous remark that the latter Lagrangian is invariant under the one-parameter
group of transformations $F\mapsto \e^{\iu\th t^1}F$ (with $\th\in\RR$) generated by the $\su(1,1)$
generator $t^1$. From the fact that Eq.~\eqref{PhiPsi} can be written in matrix form as
\[
  (\phi,\psi)^{\mathsf T}
  =\e^{-\frac{\iu\pi}4\si^y}(\Phi,\Psi)^{\mathsf T},
\]
it follows that the original Lagrangian~\eqref{L} is invariant under the one-parameter group
\[
  (\phi,\psi)^{\mathsf T}
  \mapsto \e^{-\frac{\iu\pi}4\si^y}\e^{-\th\si^x}\e^{\frac{\iu\pi}4\si^y}(\phi,\psi)^{\mathsf T}
  =\e^{\th\si^z}(\phi,\psi)^{\mathsf T},
\]
i.e., the obvious scaling symmetry
\begin{equation}\label{SU11symm}
  \phi\mapsto\e^{\th}\phi,\qquad \psi\mapsto\e^{-\th}\psi.
\end{equation}
The relation between the coupled GMTM model studied in this work and the $\SU(1,1)$ Thirring
model~\eqref{LSU11} is thus akin to the relation between the quantum XXX Heisenberg model
(invariant under the full rotation group $\mathrm{SO}(3)$) and its XXZ counterpart (only
invariant under rotations around an axis). It should be noted, in this respect, that this
residual invariance of the XXZ spin chain plays a key role in the construction of the Bethe
eigenstates of this model.

\begin{remark}
  In the quantized version of the Lagrangian~\eqref{Lalt}-\eqref{LI}, the minus sign in front of
  the second Thirring Lagrangian could perhaps be explained by regarding the $\Psi$ field as a
  kind of fermionic ghost field. With this interpretation, the quantum counterpart of~\eqref{Lalt}
  would describe the mutual interaction of a (self-interacting) Thirring fermion with a ghost
  Thirring fermion, with the same mass $m$ and opposite coupling constants $\pm g/2$. This type of
  ghost fields has recently been discussed in bosonic field theories in the context of quantized
  gravity, where the propagator of the (four-derivative) graviton can be effectively expressed as
  the sum of the standard propagators of a massless graviton field and a massive ghost field with
  negative kinetic energy~\cite{GSTZ21}.
\end{remark}

\section{Nonlocal conserved quantities}\label{sec.NCQ}

One of the key aspects in the study of the integrability of a given system is the construction of
an infinite set of conserved quantities. In this section we shall exploit the zero curvature
formulation of the equations of motion of the coupled GMTM developed above in order to obtain
several infinite hierarchies of conserved quantities. Following Ref.~\cite{Ag12}, we shall derive
these conserved quantities from the quotients
\[
  \Ga_{ij}=w_j^{-1}w_i=w_iw_j^{-1}\equiv\frac{w_i}{w_j},
\]
where $i\ne j$ and $w_j$ is assumed to be bosonic (even) and with nonvanishing body (complex
number part) to guarantee that $w_j^{-1}$ exists and is also even~\cite{De92}. From the identity
\[
  \pd_t\big(w_j^{-1}\pd_xw_j\big)=\pd_x\big(w_j^{-1}\pd_tw_j\big)
\]
and the Lax pair equations~\eqref{uvn} it then follows that
\begin{equation}
  \pd_t \Big(U_{jj}+\sum_{i;i\ne j} U_{ji}\Ga_{ij}\Big)= \pd_x \Big(V_{jj}
  +\sum_{i;i\ne j} V_{ji}\Ga_{ij}\Big).
\label{CE}
\end{equation}
If the fields vanish sufficiently fast as $|x|\to\infty$, from these equations we deduce
that\footnote{Here and in what follows, unless otherwise stated we shall assume that the
  integration range is the whole real line.}
\begin{align}
  I_j(t)= \int\diff x \Big(U_{jj}+\sum_{i;i\ne j}U_{ji}\Ga_{ij}\Big),
\label{ce}
\end{align}
is the generating function for an infinite hierarchy of conserved quantities~\cite{Ag12}. It is
also immediate to show that the auxiliary functions $\Ga_{ij}$ (with $i\ne j$) satisfy the
following pair of coupled Riccati equations:
\begin{align}
  \pd_x\Ga_{ij}&=U_{ij}-U_{jj}\Ga_{ij}+\sum_{k\ne j}\Big(U_{ik}
    -\Ga_{ij}U_{jk}\Big)\Ga_{kj},
                         \label{rcx}\\
  \pd_t\Ga_{ij}&=V_{ij}-V_{jj}\Ga_{ij}+\sum_{k\ne j}\Big(V_{ik}
                         -\Ga_{ij}V_{jk}\Big)\Ga_{kj},
  \label{rct}
\end{align}
where $U_{ij}$ and $V_{ij}$ denote the elements of the matrices $U$ and $V$. It can be readily
checked that the compatibility condition $\pd_x\pd_t\Ga_{ij}=\pd_t\pd_x\Ga_{ij}$ for the latter
system is automatically satisfied once the equations of motion for the fields $\phi$, $\psi$ are
taken into account.

In this section we shall assume that $w_1$ and $w_2$ are even (and with nonvanishing body), so
that $w_3$ is odd, and shall compute the conserved quantities obtained from the quotients
$\Ga_{i1}$ with $i=2,3$. Note that the conserved quantities constructed from $\Ga_{i2}$ with
$i=1,3$ can be derived from the former by the substitutions
\begin{equation}\label{w12map}
  \phi_1^*\leftrightarrow\psi_1,\qquad \phi_2^*\leftrightarrow -\psi_2.
\end{equation}
Indeed, under this mapping the functions in Eq.~\eqref{rhorr} transforms as
\[
  \rho_\pm\leftrightarrow-\rho_\pm,\qquad r_1^\pm\leftrightarrow r_2^\mp,
\]
which by Eq.~\eqref{u} is equivalent to reversing the roles of $w_1$ and $w_2$ in the Lax pair. In
fact, from the previous argument it follows that~\eqref{w12map} is a symmetry of the field
equations~\eqref{eqsm} themselves, and thus maps any conserved quantity of these equations into
another conserved quantity.

\subsection{Expansion in negative powers of $\la$}

The equations~\eqref{rcx}-\eqref{rct} satisfied by the quotients $\Ga_{31}$ and $\Ga_{21}$ can be
more explicitly written as the system of coupled equations
\begin{equation}
\begin{aligned}
  \pd_x\Ga_{31}&=r_2^-+r_1^+\Ga_{21}+\iu\bigg[\frac{1}{2}(\la^2-\la^{-2})
                    -\rho_-\bigg]\Ga_{31},\\
  \pd_x\Ga_{21}&=-2\iu\rho_-\Ga_{21}-r_2^+\Ga_{31}+
                    r_1^-\Ga_{21}\Ga_{31},\\[6pt]
  \pd_t\Ga_{21}&=-2\iu\rho_+\Ga_{21}-r_2^-\Ga_{31}+
                    r_1^+\Ga_{21}\Ga_{31}.
\end{aligned}
\label{rci}
\end{equation}
We have omitted the equation for $\pd_t\Ga_{31}$, since we shall see below that it is not actually
needed. In order to solve this system, we first expand $\Ga_{31}$ and $\Ga_{21}$ in negative
powers of the spectral parameter $\la$, namely
\begin{equation}
  \Ga_{i1}(x,t; \la)= \sum_{n=1}^{\infty}\la^{-n}\Ga^{(n)}_{i1}(x,t),\qquad i=2,3.
\label{1ga}
\end{equation}
The expansion coefficients $\Ga^{(n)}_{i1}$ can be computed substituting the previous
expansions in the Riccati equations~\eqref{rci}. In this way we obtain the equations
\begin{align}
  \Ga_{31}^{(n)}
  &=2\big(\psi_1\delta_{n3}-\psi_2\delta_{n1}\big) +\Ga_{31}^{(n-4)}
    +2(-\iu\pd_x+\rho_-)\Ga_{31}^{(n-2)}
    \nonumber\\
  &\qquad+2\Big(\phi^*_2\Ga_{21}^{(n-1)}+\phi^*_1\Ga_{21}^{(n-3)}\Big),
    \label{31x}\\
  \iu\,\pd_x\Ga_{21}^{(n)}
  &=2\rho_{-}\Ga_{21}^{(n)}+\psi_2\Ga_{31}^{(n+1)}
    +\psi_1\Ga_{31}^{(n-1)}\nonumber\\
  &\qquad
    +\phi^*_2\sum_{k=1}^n\Ga_{21}^{(n+1-k)}\Ga_{31}^{(k)}
    -\phi^*_1\sum_{k=1}^{n-2}\Ga_{21}^{(n-1-k)}\Ga_{31}^{(k)},    
  \label{21x}\\
  \iu\,\pd_t\Ga_{21}^{(n)}=
  &2\rho_{+}\Ga_{21}^{(n)}+\psi_2\Ga_{31}^{(n+1)}
    -\psi_1\Ga_{31}^{(n-1)}\nonumber\\
  &\qquad+\phi^*_2\sum_{k=1}^n\Ga_{21}^{(n+1-k)}\Ga_{31}^{(k)}
    +\phi^*_1\sum_{k=1}^{n-2}\Ga_{21}^{(n-1-k)}\Ga_{31}^{(k)},
\label{21t}
\end{align}
where $n\ge 1$ and $\Ga_{31}^{(n)}=\Ga_{21}^{(n)}=0$ for $n\le0$. The previous equations can
be recursively solved as follows. To begin with, Eq.~\eqref{31x} with $n=1,2$ yields
\begin{equation}\label{Ga3112}
  \Ga_{31}^{(1)}=-2\psi_2,\qquad \Ga_{31}^{(2)}=2\phi_2^*\Ga_{21}^{(1)}.
\end{equation}
Substituting the previous expressions for $\Ga_{31}^{(1)}$ and $\Ga_{31}^{(2)}$ into
Eq.~\eqref{21x} we then obtain a first-order homogeneous equation for $\Ga_{21}^{(1)}$, namely
\[
  \pd_x\Ga_{21}^{(1)}=-2\iu\rho_-\Ga_{21}^{(1)}-2\iu\psi_2\phi^*_2\Ga_{21}^{(1)}
  +2\iu\phi_2^*\psi_2\Ga_{21}^{(1)}=2\iu\rho_+\Ga_{21}^{(1)}\,.
\]
Integrating this equation we obtain
\begin{equation}\label{Ga211}
  \Ga_{21}^{(1)}=A_1R(x,t),
\end{equation}
where
\begin{equation}
  \label{Rdef}
  R(x,t):=\e^{2\iu\int_{-\infty}^x\diff y\,\rho_+(y,t)}
\end{equation}
and $A_1$ is a bosonic function of $t$ alone. We still must enforce Eq.~\eqref{21t} with $n=1$,
which reads
\[
  \pd_t\Ga_{21}^{(1)}=2\iu\rho_-\Ga_{21}^{(1)}.
\]
Using the explicit expression~\eqref{Ga211} for $\Ga_{21}^{(1)}$ and the relation
\begin{equation}\label{rhoxt}
  \pd_t\rho_+=\pd_x\rho_-,
\end{equation}
which easily follows from the equations of motion, we obtain (taking into account that $A_1$ is
even and denoting the derivative with respect to $t$ by a dot)
\[
  R(x,t)^{-1}\pd_t\Ga_{(21)}^{(1)}=2\iu\rho_-A_1
  =\dot A_1+2\iu A_1\int_{-\infty}^x\diff y\,\pd_t\rho_+(y,t)
  =  \dot A_1+2\iu A_1\rho_-.
\]
Thus $A_1$ is a constant. This determines $\Ga_{21}^{(1)}$, which yields $\Ga_{31}^{(2)}$ through
Eq.~\eqref{Ga3112}.

In general, assume that $\Ga_{21}^{(k)}$ (with $k\le n-1$) and $\Ga_{31}^{(k)}$ (with $k\le n$)
have been computed. Combining Eqs.~\eqref{31x} (with $n+1$ instead of $n$) and
\eqref{21x}-\eqref{21t} we easily obtain a linear inhomogeneous system of the form
\begin{equation}\label{Ga21n}
  \pd_x\Ga_{21}^{(n)}=2\iu\rho_+\Ga_{21}^{(n)}+B_n,\qquad
  \pd_t\Ga_{21}^{(n)}=2\iu\rho_-\Ga_{21}^{(n)}+C_n,
\end{equation}
where $B_n$ and $C_n$ depend on the \emph{known} functions $\Ga_{21}^{(1)},\dots,\Ga_{21}^{(n-1)}$
and $\Ga_{31}^{(1)},\dots,\Ga_{31}^{(n)}$ by the induction hypothesis. Note that these equations
are automatically compatible, due to the compatibility of Eqs.~\eqref{rcx}-\eqref{rct}. From
Eq.~\eqref{rhoxt} and the compatibility condition for the system~\eqref{Ga21n} we arrive at the
identity
\begin{equation}
  \label{compGa21n}
  (\pd_x-2\iu\rho_+)C_n=(\pd_t-2\iu\rho_-)B_n.
\end{equation}
Integrating the first equation in~\eqref{Ga21n} with respect to $x$ we easily obtain:
\begin{equation}\label{Ga21nint}
  \Ga_{21}^{(n)}= R(x,t)\left( A_{n}(t)+\int_{-\infty}^x\diff y\,R(y,t)^{-1}B_n(y,t)\right).
\end{equation}
The (bosonic) function $A_{n}(t)$ is then determined (up to an arbitrary constant) substituting
this expression into the second Eq.~\eqref{Ga21n}, which yields the differential equation
\begin{equation}
  \label{dotAn}
  \dot A_n(t)=R^{-1}C_n-\int_{-\infty}^x\diff
  y\,R^{-1}(y,t)\Big(\pd_t-2\iu\rho_-(y,t)\Big)B_n(y,t).
\end{equation}
Note that the right-hand side (RHS) of this equation is indeed a function of $t$ alone, since its
derivative with respect to $x$,
\[
  R^{-1}\Big[(-2\iu\rho_+C_n+\pd_xC_n-(\pd_t-2\iu\rho_-)B_n\Big],
\]
vanishes on account of Eq.~\eqref{compGa21n}. In this way we can determine $\Ga_{21}^{(n)}$, which
in turn yields $\Ga_{31}^{(n+1)}$ through Eq.~\eqref{31x} with $n$ replaced by $n+1$.

By Eq.~\eqref{ce} with $j=1$, the generating function for the conserved quantities constructed from
$\Ga_{31}$ and $\Ga_{21}$ reads
\begin{align}
  I_1=\int\diff x \Big(U_{11}+U_{13}\Ga_{31}\Big)=\iu\int\diff x \Big[\rho_-
  +\Big(\la\phi^*_2-\la^{-1}\phi^*_1\Big)\Ga_{31}\Big],
\label{ce1}
\end{align}
where we have discarded the trivial constant term in $U_{11}$. Substituting the
expansion~\eqref{1ga} of $\Ga_{31}$ into the above equation we obtain a corresponding expansion
\[
  I_1=\iu\sum_{n=0}^\infty\la^{-n}I_1^{(n)},
\]
where each coefficient $I_1^{(n)}$ is a conserved quantity. The general expression for these
conserved quantities is given by
\begin{align}
  I_{1}^{(n)}=\int\diff
  x \Big(\rho_{-}\delta_{n0}+\phi^*_{2}\Ga_{31}^{(n+1)}-\phi^*_{1}\Ga_{31}^{(n-1)}\Big),\qquad
  n=0,1,\dots,.
\label{cq1}
\end{align}

The explicit form of the conserved quantities~\eqref{cq1} can be obtained recursively from
Eq.~\eqref{31x}. The first two nontrivial conserved quantities turn out to be local, and are given
by
\begin{align}
  I_1^{(0)}&= -\int\diff x\, (\phi^*_1\psi_1 + \phi^*_2\psi_2 ),\\
  I_1^{(2)}&=2\int\diff x\,\Big(2\iu\,\phi^*_2\psi_{2,x}+ \phi^*_1 \psi_2 + \phi^*_2\psi_1 
             + 2\phi^*_1 \psi_1\phi^*_2\psi_2\Big),
\end{align}
while $I_1^{(1)}$ vanishes identically. On the other hand, the conserved quantities $I_1^{(n)}$
with $n\ge 3$ are nonlocal. For instance, from Eq.~\eqref{cq1} with $n=3$ and the expressions for
$\Ga_{31}^{2}$, $\Ga_{31}^{(4)}$ we readily obtain
\begin{equation}\label{I13}
  I_1^{(3)}=4\int\diff x\,\phi_2^*\Big(\phi^*_1-\iu\phi^*_{2,x}\Big)\Ga_{21}^{(1)}=
  4A_1\int\diff x\,\phi_2^*\Big(\phi^*_1-\iu\phi^*_{2,x}\Big)R(x,t),
\end{equation}
where $A_1$ is a bosonic constant and $R(x,t)$ is defined by Eq.~\eqref{Rdef}
(cf.~Eq.~\eqref{Ga211}). Proceeding in the same way, a long but straightforward calculation yields
\begin{align}
  I_1^{(4)}&=-2\int\diff x\Big(2\iu\big(\phi_1^*\psi_{2,x}+\phi_2^*\psi_{1,x}\big)
             -4\phi_2^*\psi_{2,xx}+4\iu\,
             \phi_1^*\phi_{2}^*\big(\psi_2\psi_{1,x}-2\psi_1\psi_{2,x}\big)\nonumber\\
           &\quad+4\iu\phi_2^*\psi_1\psi_2\phi_{1,x}^*
             +\rho_{+}+2\phi^*_2\big(\iu\phi_{2,x}^*-\phi^*_{1}\big)\Ga^{(2)}_{21}\Big).
             \label{I14}
\end{align}
The bosonic function $\Ga^{(2)}_{21}$ is obtained integrating Eqs.~\eqref{21x}-\eqref{21t} with
$n=2$, which in this case can be seen to reduce to the system~\eqref{Ga21n} with
\[
  B_2=4\psi_2(\psi_{2,x}-\iu\psi_1),\qquad C_2=4\psi_2\psi_{2,x}.
\]
From Eq.~\eqref{Ga21nint} with $n=2$ we then obtain
\begin{align}
  \Ga_{21}^{(2)}=R(x)\bigg[A_2(t)+4\int_{-\infty}^x\diff y\,
  R(y)^{-1}\psi_2(y)\Big(\psi_{2,x}(y)-\iu\,\psi_1(y)\Big)\bigg],
\end{align}
where for the sake of simplicity, we have dropped the $t$ dependence of the fields and the
function $R$. The bosonic function $A_2(t)$ is determined (up to a constant) by Eq.~\eqref{dotAn}
with $n=2$, namely
\begin{equation}\label{dotA2}
  \dot A_2(t)=4R^{-1}\psi_2\psi_{2,x}-4\int_{-\infty}^x\diff
  y\,R^{-1}(y)\Big(\pd_t-2\iu\rho_-(y)\Big)
  \Big(\psi_2(y)\psi_{2,x}(y)+\iu\,\psi_1(y)\psi_2(y)\Big).
\end{equation}
As explained above, it is guaranteed that the RHS of the latter equation is independent of $x$.
This can also be explicitly checked by noting that the $x$ derivative of the RHS of
Eq.~\eqref{dotA2} multiplied by $R(x,t)/4$ is explicitly given by
\begin{multline*}
  -2\iu\,\rho_+\psi_2\psi_{2,x}+\psi_2\psi_{2,xx}-(\pd_t-2\iu\,\rho_-)
  (\psi_2\psi_{2,x}+\iu\psi_1\psi_2)\\
  =4\iu\,\phi_1^*\psi_1\psi_2\psi_{2,x}+\psi_2\psi_{2,xx}-\pd_t(\psi_2\psi_{2,x}+\iu\psi_1\psi_2),
\end{multline*}
which is readily seen to vanish using the equations of motion of the fields.

By the remark at the beginning of this section, the conserved quantities $I_2^{(n)}$ constructed
from $\Ga_{i2}$ with $i\ne2$ can be derived from the ones obtained above through the
mapping~\eqref{w12map}. In this way we obtain the explicit formulas
\begin{align}
  I_2^{(0)}&= \int\diff x\, (\phi^*_1\psi_1 + \phi^*_2\psi_2 ),\\
  I_2^{(2)}&=2\int\diff x\,\Big(2\iu\,\psi_2\phi^*_{2,x}+ \phi^*_1 \psi_2 + \phi^*_2\psi_1 
             + 2\phi^*_1 \psi_1\phi^*_2\psi_2\Big),\\
  I_2^{(3)}&=-4A_1\int\diff x\,\psi_2\Big(\psi_1+\iu\psi_{2,x}\Big)R(x,t)^{-1},\\
  I_2^{(4)}&=-2\int\diff x\Big(2\iu\big(-\psi_1\phi^*_{2,x}-\psi_2\phi^*_{1,x}\big)
             -4\psi_2\phi^*_{2,xx}+4\iu\,\psi_1\psi_{2}\big(\phi_2^*\phi^*_{1,x}
             -2\phi_1^*\phi^*_{2,x}\big)\nonumber\\
           &\quad+4\iu\phi^*_1\phi^*_2\psi_2\psi_{1,x}
             -\rho_{+}+2\psi_2\big(\iu\psi_{2,x}+\psi_{1}\big)\Ga^{(2)}_{12}\Big),
\end{align}
where $A_1$ is again an arbitrary bosonic constant,
\begin{align}
  \Ga_{12}^{(2)}=R(x)^{-1}\bigg[A_2(t)+4\int_{-\infty}^x\diff y\,
  R(y)\phi_2^*(y)\Big(\phi^*_{2,x}(y)+\iu\,\phi^*_1(y)\Big)\bigg],
\end{align}
and $A_2(t)$ is a bosonic function determined (up to a constant) by the equation
\begin{equation}\label{2dotA2}
  \dot A_2(t)=4R\phi^*_2\phi^*_{2,x}-4\int_{-\infty}^x\diff
  y\,R(y)\Big(\pd_t+2\iu\rho_-(y)\Big)
  \Big(\phi_2^*(y)\phi^*_{2,x}(y)-\iu\,\phi^*_1(y)\phi^*_2(y)\Big).
\end{equation}

\subsection{Expansion in positive powers of $\la$}
A second hierarchy of conserved quantities can be obtained by expanding $\Ga_{i1}$ with $i\ne 1$
in positive powers of $\la$ as follows:
\begin{align}
\Ga_{i1}(x,t; \la)=\sum_{k=1}^{\infty}\wt{\Ga}^{(k)}_{i1}(x,t)\la^{k},\qquad 1=2,3.
\label{2ga}
\end{align}
Using the procedure described above, the expansion coefficients $\wt{\Ga}^{(k)}_{i1}$ can
easily be obtained in a recursive way from the Riccati equations~\eqref{rci}, which yield system
\begin{align}
  \wt\Ga_{31}^{(n)}
  &=2\big(\psi_2\delta_{n3}-\psi_1\delta_{n1}\big) +\wt\Ga_{31}^{(n-4)}
    +2(\iu\pd_x-\rho_-)\wt\Ga_{31}^{(n-2)}
    \nonumber\\
  &\qquad-2\Big(\phi^*_2\wt\Ga_{21}^{(n-3)}+\phi^*_1\wt\Ga_{21}^{(n-1)}\Big),
    \label{31p}\\
  \iu\,\pd_x\wt\Ga_{21}^{(n)}
  &=2\rho_{-}\wt\Ga_{21}^{(n)}+\psi_2\wt\Ga_{31}^{(n-1)}
    +\psi_1\wt\Ga_{31}^{(n+1)}\nonumber\\
  &\qquad
    +\phi^*_2\sum_{k=1}^{n-2}\wt\Ga_{21}^{(n-1-k)}\wt\Ga_{31}^{(k)}
    -\phi^*_1\sum_{k=1}^{n}\wt\Ga_{21}^{(n+1-k)}\wt\Ga_{31}^{(k)},    
  \label{21p}\\
  \iu\,\pd_t\wt\Ga_{21}^{(n)}
  &=2\rho_{+}\wt\Ga_{21}^{(n)}+\psi_2\wt\Ga_{31}^{(n-1)}
    -\psi_1\wt\Ga_{31}^{(n+1)}\nonumber\\
  &\qquad
    +\phi^*_2\sum_{k=1}^{n-2}\wt\Ga_{21}^{(n-1-k)}\wt\Ga_{31}^{(k)}
    +\phi^*_1\sum_{k=1}^{n}\wt\Ga_{21}^{(n+1-k)}\wt\Ga_{31}^{(k)},
\label{21pt}
\end{align}
where $n\ge 1$ and $\wt{\Ga}_{31}^{(n)}=\wt{\Ga}_{21}^{(n)}=0$ for $n\le0$. Substituting the
expansion~\eqref{2ga} of $\wt{\Ga}_{31}$ into Eq.~\eqref{ce1} we obtain an expansion of the form
\[
  I_1= \iu\,\sum_{n=0}^{\infty}\la^n\wt I_1^{(n)},
\]
whose coefficients
\begin{equation}\label{cq2}
  \wt{I}_{1}^{(n)}=\int\diff x\Big(\rho_{-}\delta_{n0}+\phi^*_{2}\wt{\Ga}_{31}^{(n-1)}
  -\phi^*_{1}\wt{\Ga}_{31}^{(n+1)}\big),\qquad n=0,1,\dots,
\end{equation}
are conserved quantities. The explicit form of the first few nontrivial conserved quantities is as
follows:
\begin{align}
  \wt{I}_1^{(0)}&= \int\diff x\,\big(\phi^*_1\psi_1 + \phi^*_2\psi_2\big), \nonumber \\
  \wt{I}_1^{(2)}&=2 \int\diff x\,\Big(2\iu \phi^*_1 \psi_{1x}  - \phi^*_1\psi_2- \phi^*_2 \psi_1 
                  -2\phi^*_1\psi_1\phi^*_2\psi_2\Big) \nonumber,\\
  \wt{I}_1^{(3)}&=4\wt{A}_1 \int\diff
                  x\,\phi^*_1\big(\phi_2^*+\iu\phi^*_{1x}\big)R(x,t)^{-1},
                  \label{tI1}
\end{align}
where $\wt{A}_1$ is a bosonic constant. As was the case with the conserved quantities $I_1^{(n)}$
discussed above, all the conserved quantities $\wt I_1^{(n)}$ with $n\ge3$ are nonlocal. The
corresponding conserved quantities constructed from $\Ga_{i2}$ with $i\ne2$ are obtained from the
above through the mapping~\eqref{w12map}, namely
\begin{align}
  \wt{I}_2^{(0)}&= -\int\diff x\,\big(\phi^*_1\psi_1 + \phi^*_2\psi_2\big), \nonumber \\
  \wt{I}_2^{(2)}&=2 \int\diff x\,\Big(2\iu \psi_1 \phi^*_{1x} +\phi^*_1\psi_2+\phi^*_2 \psi_1 
                  -2\phi^*_1\psi_1\phi^*_2\psi_2\Big) \nonumber,\\
  \wt{I}_2^{(3)}&=4\wt{A}_1 \int\diff
                  x\,\psi_1\big(\psi_2+\iu\psi_{1x}\big)R(x,t),
                  \label{tI2}
\end{align}
where again $\wt A_1$ is an arbitrary bosonic constant.
\begin{remark}
  As is also the case with the ordinary GMTM~\cite{Ag12}, the conserved quantities $I^{(k)}_j$,
  $\wt{I}^{(k)}_j$ with $k\le2$ are local, while the ones with $k\ge 3$ are nonlocal. Note,
  however, that all the conserved quantities of the coupled BMTM introduced in Ref.~\cite{BS23}
  are local.
\end{remark}
\begin{remark}
  There are some obvious relations among the lowest-order conserved quantities derived above,
  namely
  \[
    I_1^{(0)}=\wt{I}_2^{(0)}=-\wt{I}_1^{(0)}=-I_2^{(0)},\qquad
    I_1^{(2)}=I_2^{(2)},\qquad \wt{I}_1^{(2)}=\wt{I}_2^{(2)},
  \]
  where for the last two equalities we have taken into account that the integral of a total $x$
  derivative vanishes due to the boundary conditions imposed on the fields. Note, however, that no
  such relations are apparent for the conserved quantities of order greater than $2$.
\end{remark}
\begin{remark}
  Setting
  \[
    c=\frac{\Ga_{31}}{\iu\la},\qquad b=\Ga_{21},
  \]
  the Riccati system~\eqref{rci} can be rewritten as follows:
  \begin{align*}
    \pd_x c&= \psi_2-\la^{-2}\psi_1-(\phi^*_2+\la^2\phi^*_1)bc,\\
    \pd_x b&= -2\iu\rho_-b+(\la^2\psi_2+\psi_1)c+(\la^2\phi^*_2-\phi^*_1)bc,\\
    \pd_t b&= -2\iu\rho_+b+(\la^2\psi_2-\psi_1)c+(\la^2\phi^*_2+\phi^*_1)bc.
  \end{align*}
  Since the RHS of the latter equations depends on $\la$ only through $\la^2$, it is clear that it
  admits solutions in which $c$ and $b$ are functions of $\la^2$. Hence the original
  equations~\eqref{rci} also admit solutions in which $\Ga_{31}$ is odd in $\la$ and $\Ga_{21}$ is
  even in $\la$. By Eqs~\eqref{cq1}-\eqref{cq2}, the conserved quantities $I_1^{(n)}$ and
  $\wt I_1^{(n)}$ constructed from these solutions vanish identically when $n$ is odd. Applying
  the symmetry transformation~\eqref{w12map} we deduce that the same is true for the functions
  $\Ga_{i2}$ with $i\ne2$ and their corresponding conserved quantities~$I_2^{(n)}$,
  $\wt I_2^{(n)}$ of odd order.
\end{remark}

\section{Local conserved quantities}\label{sec.LCQ}

In this section we shall regard the auxiliary fields $w_{1,2}$ as fermionic and $w_3$ as bosonic,
and construct the conserved quantities derived from the quotients $\Ga_{i3}$ with $i\ne3$.
According to the general equation~\eqref{ce}, the generating function for these quantities is
\begin{equation}
  \label{I3t}
  I_3=\int\diff x\Big(U_{31}\Ga_{13}+U_{32}\Ga_{23}\Big)=\iu\int\diff x\Big[(\la\psi_2
  -\la^{-1}\psi_1)\Ga_{13}-(\la\phi_2^*+\la^{-1}\phi_1^*)\Ga_{23}\Big],
\end{equation}
where we have again dropped the trivial constant term $U_{33}$. The differential equations for the
functions $\Ga_{i3}$ (with $i\ne3$) read
\begin{align*}
  \pd_x\Ga_{13}&=-r_1^-+\iu\bigg[\rho_--\frac12(\la^2-\la^{-2})\bigg]\Ga_{13}
                 +r_1^+\Ga_{13}\Ga_{23},\\
  \pd_t\Ga_{13}&=-r_1^++\iu\bigg[\rho_+-\frac12(\la^2+\la^{-2})\bigg]\Ga_{13}
                 +r_1^-\Ga_{13}\Ga_{23},\\
  \pd_x\Ga_{23}&=-r_2^+-\iu\bigg[\rho_-+\frac12(\la^2-\la^{-2})\bigg]\Ga_{23}
                 -r_2^-\Ga_{13}\Ga_{23},\\
  \pd_t\Ga_{23}&=-r_2^--\iu\bigg[\rho_++\frac12(\la^2+\la^{-2})\bigg]\Ga_{23}
                 -r_2^+\Ga_{13}\Ga_{23}.
\end{align*}
The structure of these equations makes it convenient to introduce light-cone coordinates
\[
  \xi=\frac12(t+x),\qquad \eta=\frac12(t-x),
\]
so that
\[
  \pd_\xi=\pd_t+\pd_x,\qquad \pd_\eta=\pd_t-\pd_x\,.
\]
Setting
\[
  \frac{\Ga_{13}}{2\la}=c\,,\qquad \frac{\Ga_{23}}{2\la}=b%,\qquad\mu=\iu\la^2
\]
we obtain the system
\begin{align}
  \pd_\xi c&=\iu\phi^*_2+(2\iu\rho_2-\mu)c+4\mu\phi^*_2bc,\label{xic}\\
    \pd_\xi b&=-\iu\psi_2-(2\iu\rho_2+\mu)b+4\mu\psi_2bc,\label{xib}\\
  \pd_\eta c&=-\mu^{-1}\phi^*_1+(2\iu\rho_1+\mu^{-1})c-4\iu\phi^*_1bc,\label{etac}\\
  \pd_\eta b&=-\mu^{-1}\psi_1-(2\iu\rho_1-\mu^{-1})b+4\iu\psi_1bc,\label{etab}
\end{align}
where $\mu=\iu\la^2$ is a new (complex) spectral parameter and
\[
  \rho_i=\phi^*_i\psi_i\,,\qquad i=1,2\,.
\]
Using the field equations~\eqref{eqsm}, it can be readily checked that the compatibility
conditions for the previous system are automatically verified. The main difference with the
procedure followed in the previous section is that, as we shall show below, when the functions $c$
and $b$ are expanded in powers of the spectral parameter $\mu$ either Eqs.~\eqref{xic}-\eqref{xib}
(when expanding in negative powers) or \eqref{etac}-\eqref{etab} (when expanding in positive
powers) can be used to generate a pure recursion relation for the expansion coefficients of the
latter functions. In this way these coefficients can be recursively determined in terms of the
fields without having to perform any integrations.
\begin{remark}
  Equations~\eqref{xic}-\eqref{etac} are easily seen to be invariant under the transformation
  \[
    c\leftrightarrow b^*,\qquad \phi_i\leftrightarrow\psi_i.
  \]
  Since our model reduces to two identical copies of the original Thirring mode when $\phi=\psi$,
  it follows that (with suitable boundary conditions) one can take $b=c^*$ in the latter model.
  This simplification, actually used in Ref.~\cite{IK78}, is of course impossible in our case.
\end{remark}
\subsection{Negative powers of $\mu$}

Let us start by expanding the (fermionic) functions $c$ and $b$ in negative powers of the spectral
parameter $\mu$, namely
\[
  c=\sum_{n\ge1}c_n\mu^{-n},\qquad b=\sum_{n\ge1}b_n\mu^{-n}.
\]
Substituting into Eqs.~\eqref{xic}-\eqref{xib} we immediately obtain the coupled recursion
relations
\begin{align}
  c_{n+1}&=\iu\phi^*_2\de_{n0}+(2\iu\rho_2-\pd_\xi)c_n+4\phi^*_2\sum_{k=1}^nb_{n+1-k}c_k,
  \label{cnp1}\\
  b_{n+1}&=-\iu\psi_2\de_{n0}-(2\iu\rho_2+\pd_\xi)b_n+4\psi_2\sum_{k=1}^nb_{n+1-k}c_k
  \label{bnp1}
\end{align}
with $n=0,1,\dots$. Since the RHS of these equations contain only functions $c_k$ and $b_k$ with
$1\le k\le n$, it is obvious that they allow the recursive computation of $c_n$ and $b_n$ for all
$n\ge 1$. Once this is done, the corresponding conserved quantities $I_3^{(n)}$ are obtained 
expanding the generating function~\eqref{I3t}, which can be written as
\begin{equation}\label{I3}
  I_3=2\int\diff x\,\Big[(\mu\psi_2-\iu\psi_1)c-(\mu\phi^*_2+\iu\phi^*_1)b\Big]
\end{equation}
in powers of $\mu$. Setting
\[
  I_3=2\sum_{n\ge0} \mu^{-n}I_3^{(n)}
\]
we thus obtain the explicit expression
\begin{equation}
  \label{I3n}
  I_3^{(n)}=\int\diff x\,\Big[\psi_2c_{n+1}-\phi^*_2b_{n+1}-\iu(\psi_1c_n+\phi^*_1b_n)\Big],\qquad
  n\ge0\,.
\end{equation}
From the latter expression it is obvious that these conserved quantities are all local. The first
of these quantities, $I_3^{(0)}$, is easily seen to vanish, since from
Eqs.~\eqref{cnp1}-\eqref{bnp1} we have
\[
  c_1=\iu\phi^*_2,\qquad b_1=-\iu\psi_2\,.
\]
The first nontrivial conserved density is easily obtained setting $n=1$ in the recursion
relations~\eqref{cnp1}-\eqref{bnp1}, which yields
\[
  c_2=(2\iu\rho_2-\pd_\xi)c_1+4\phi^*_2b_1c_1=-\iu(2\iu\rho_2+\pd_\xi)\phi^*_2=-\iu\phi^*_{2,\xi}.
\]
Although $b_2$ can be computed in a similar way from Eq.~\eqref{bnp1} with $n=1$, it is easier to
note that $c=\Ga_{13}/(2\la)$ is mapped into $b=\Ga_{23}/(2\la)$ under the symmetry
transformation~\eqref{w12map}. In this way we obtain
\[
  b_2=\iu\psi_{2,\xi},
\]
and from Eq.~\eqref{I3n} with $n=1$ we have
\[
  I_3^{(1)}=-\int\diff
  x\Big[\iu(\psi_2\phi^*_{2,\xi}+\phi^*_2\psi_{2,\xi})+\phi^*_1\psi_2+\phi^*_2\psi_1\Big].
\]
Using the field equations~\eqref{eqsm} we can express the $\xi$ derivatives of $\psi_2$ and
$\phi^*_2$ in terms of their $x$ derivatives as follows:
\[
  \psi_{2,\xi}=2\psi_{2,x}-\iu\psi_1-2\iu\rho_1\psi_2,\qquad
  \phi^*_{2,\xi}=2\phi^*_{2,x}+\iu\phi^*_1+2\iu\rho_1\phi^*_2.
\]
Substituting into the previous expression for $I_3^{(1)}$ we finally obtain the explicit formula
\begin{equation}
  \label{I311}
  I_3^{(1)}=-2\int\diff x\Big[\iu\Big(\phi_2^*\psi_{2,x}-\phi^*_{2,x}\psi_2\Big)+\phi^*_1\psi_2+\phi^*_2\psi_1+2\rho_1\rho_2\Big].
\end{equation}
Integrating by parts we obtain the equivalent expression
\begin{equation}
  \label{I3112}
  I_3^{(1)}=-2\int\diff
  x\Big(2\iu\phi_2^*\psi_{2,x}+\phi^*_1\psi_2+\phi^*_2\psi_1+2\rho_1\rho_2\Big).
%  =-I_1^{(2)}
\end{equation}
% (cf.~Eq.~\eqref{}).
Similarly, from Eqs.~\eqref{cnp1}-\eqref{bnp1} with $n=2$ we obtain
\[
  c_3=\iu\phi^*_{2,\xi\xi}-2\rho_2\phi^*_{2,\xi},\qquad
  b_3=-\iu\psi_{2,\xi\xi}-2\rho_2\psi_{2,\xi}.
\]
However, the corresponding conserved quantity $I_3^{(2)}$ is trivial, since using the field
equations it can be shown that
\[
  I_3^{(2)} =2\int\diff
  x\,\pd_x\Big(4\iu\phi_2^*\psi_{2,x}+\phi^*_1\psi_2+\phi^*_2\psi_1+4\rho_1\rho_2\Big) =0
\]
is the integral of a total $x$ derivative, which vanishes on account of the boundary conditions at
infinity.

The calculation of the conserved quantities of order higher than 2 becomes increasingly more
involved. For example, for $n=3$ we have
\[
  c_4=-\iu\phi^*_{2,\xi\xi\xi}+4\rho_2\phi^*_{2,\xi\xi}-6\phi^*_2\phi^*_{2,\xi}\psi_{2,\xi},\qquad
  b_4=\iu\psi_{2,\xi\xi\xi}+4\rho_2\psi_{2,\xi\xi}+6\psi_2\psi_{2,\xi}\phi^*_{2,\xi},
\]
and hence
\begin{multline*}
  I_3^{(3)}=\int\diff x\,\Big(-\iu
  \phi_2^*\psi_{2,\xi\xi\xi}-\iu \psi_2\phi_{2,\xi\xi\xi}^*-\phi_1^*\psi_{2,\xi\xi}+\psi
  _1\phi_{2,\xi\xi}^{* }\\
  +2\rho_2(\iu \phi_1^*\psi_{2,\xi}+\iu
  \psi_1\phi_{2,\xi}^{* }+6\phi_{2,\xi}^*\psi_{2,\xi})\Big),
\end{multline*}
where the $\xi$ derivatives must be expressed in terms of $x$ derivatives using the field
equations~\eqref{eqsm}. When this is done we obtain the explicit expression
\begin{align*}
  I_3^{(3)}={}&2\int
  \diff x\Big[
    4 \phi_{1,x}^*\psi_{2,x}+4\phi_{2,x}^*\psi_{1,x}-8\phi_1^*\phi_2^*\psi_{1,x}\psi_{2,x}
    +24(\rho_1+\rho_2)\phi_{2,x}^*\psi_{2,x}\\
  &\en-8\phi_1^*\psi_2\phi_{2,x}^*\psi_{1,x}
    -8\phi_2^*\psi_1\phi_{1,x}^*\psi_{2,x}
    -8 \psi_1\psi_2\phi_{1,x}^*\phi_{2,x}^*+\phi_1^*\psi_2+\phi_2^*\psi_1
    +18 \rho_1\rho_2\\
  &\en+2\iu\Big(
    4\phi_{2,x}^*\psi_{2,xx}+\phi_1^*\psi_{1,x}+2\phi_2^*\psi_{2,x}
    +2\rho_1(\psi_2\phi_{1,x}^*+\phi_2^*\psi_{1,x})
    +10\rho_2(\phi_1^*\psi_{2,x}+\psi_1\phi_{2,x}^*)\Big)
    \Big],
\end{align*}
where we have discarded a total derivative with respect to $x$ in the integrand. We have verified
with \emph{Mathematica}\texttrademark{} that the even-order conserved quantities $I_3^{(4)}$ and
$I_3^{(6)}$ are trivial, as their densities are a total $x$ derivative. In fact, we conjecture
that this is the case for all conserved quantities of even order.

\subsection{Positive powers of $\mu$}

We shall next expand the fermionic functions $c$ and $b$ in positive powers of the spectral
parameter $\mu$, namely
\[
  c=\sum_{n\ge0}\mu^n\tc_n,\qquad b=\sum_{n\ge0}\mu^n\tb_n,
\]
where the inclusion of the term independent of $\mu$ is justified by the structure of
Eqs.~\eqref{etac}-\eqref{etab}. Substituting this expansion into the latter equations we obtain
the coupled recursion relation
\begin{align}
  \tc_{n+1}&=\phi^*_1\de_{n,-1}+(\pd_\eta-2\iu\rho_1)\tc_n+4\iu\phi^*_1\sum_{k=0}^n\tb_{n-k}\tc_k,
  \label{hcnp1}\\
  \tb_{n+1}&=\psi_1\de_{n,-1}+(\pd_\eta+2\iu\rho_1)\tb_n-4\iu\psi_1\sum_{k=0}^n\tb_{n-k}\tc_k.
  \label{hbnp1}
\end{align}
Likewise, from the expansion
\[
  I_3=2\sum_{n\ge0} \mu^{n}\wt I_3^{(n)}
\]
and Eq.~\eqref{I3} it follows that
\begin{equation}
  \label{tI3n}
  \wt I_3^{(n)}=\int\diff x\Big[\psi_2c_{n-1}-\phi^*_2b_{n-1}
  -\iu(\psi_1c_{n}+\phi^*_1b_{n})\Big],\qquad
  n\ge0\,.
\end{equation}
The calculation of the expansion coefficients~$\tc_n$, $\tc_n$ from the recursion
relations~\eqref{hcnp1}-\eqref{hbnp1} and the corresponding first integrals $\wt I_3^{(n)}$
from~\eqref{tI3n} proceeds along the same line as in the previous subsection. The first nontrivial
conserved quantities are
\begin{align}
  \wt I_3^{(1)}
  &=-2\int\diff x\,\Big({-}\iu \phi_1^*\psi_{1,x}-\iu\psi_1\phi_{1,x}^*+\phi_1^*\psi
    _2+\phi_2^*\psi_1+2\rho_1\rho_2\Big)\notag\\
  &= -2\int\diff x\,\Big(-2\iu \phi_1^*\psi
    _{1,x}+\phi_1^*\psi_2+\phi_2^*\psi_1+2\rho_1\rho_2\Big),  \label{tI31}\\
  \wt I_3^{(3)}&=2\int\diff x\,\Big[
                 8\rho_2(\psi_1\phi_{1,xx}^*-\phi_1^*\psi_{1,xx}+\phi_{1,x}^*\psi_{1,x})
                 +24\rho_1\phi_{1,x}^*\psi_{1,x}+4 \phi_{1,x}^*\psi_{2,x}+4\phi_{2,x}^*\psi_{1,x}
  \notag\\
  &\hphantom{2\int\en}+\phi_1^*\psi_2
    +\phi_2^*\psi_1+18\rho_1\rho_2
    -2\iu\Big(
    4 \phi_{1,x}^*\psi_{1,xx}+2 \psi_1\phi_{1,x}^*+\psi_2\phi_{2,x}^*\notag\\
  &\hphantom{2\int\qquad}+10 \rho_1(\phi_2^*\psi _{1,x}+\psi_2\phi_{1,x}^*)
    +2\rho_2(\phi_1^*\psi _{2,x}
    +\psi_1\phi_{2,x}^*)
    \Big)
    \Big],
    \label{tI33}
    \end{align}
where in the last equation we have discarded total $x$ derivatives in the integrand to simplify
the expression for the corresponding conserved quantity. As before, the conserved quantity
$\wt I_3^{(0)}$ is identically zero, while $\wt I_3^{(2)}$ is easily seen to vanish on account of
the boundary conditions at spatial infinity:
\[
  I_3^{(2)}=2\int\diff x\,\pd_x\Big({-}4\iu \phi_1^*\psi _{1,x}+\phi_1^*\psi
  _2+\phi_2^*\psi _1+4\rho_1\rho_2\Big)=0.
\]
We again conjecture that all the even-order integrals $\wt I_3^{(2n)}$ with $n=0,1,\dots$ are
trivial; in fact, we have verified this conjecture for the additional cases $n=2,3$ with the help
of \emph{Mathematica}\texttrademark{}.
\begin{remark}
  The conserved quantities of order $1$ are obviously related to the local conserved quantities
  obtained in the previous section, namely
  \[
    I_3^{(1)}=-I_{1,2}^{(2)},\qquad \wt I_3^{(1)}=\wt I_{1,2}^{(2)}.
  \]
  Note, however, that the zeroth order local conserved quantity~$\int\diff x\,\rho_+$ does not
  appear among the conserved quantities obtained in this section.
\end{remark}
\begin{remark}
  None of the conserved quantities computed above are real. Note, however, that if $I$ is a
  conserved quantity so is its complex conjugate $I^*$. It follows that the complex conjugates of
  the quantities $I_i^{(n)}$, $\wt I_i^{(n)}$ (with $i=1,2,3$) are also conserved, and can be used
  to obtain the real conserved quantities
  \[
    \Re I_i^{(n)}=\frac12\Big(I_i^{(n)}+(I_i^{(n)})^*\Big),\quad
    \Im I_i^{(n)}=\frac1{2\iu}\Big(I_i^{(n)}-(I_i^{(n)})^*\Big),
  \]
  and similarly for $\wt I_i^{(n)}$. For instance, from the complex conserved quantity $\int\diff
  x\,\rho_+$ we obtained the real conserved quantities
  \[
    \frac12\int\diff x\,\Big(\phi_1^*\psi_1+\phi_2^*\psi_2+\psi_1^*\phi_1+\psi_2^*\phi_2),\qquad
    \frac1{2\iu}\int\diff x\,\Big(\phi_1^*\psi_1+\phi_2^*\psi_2-\psi_1^*\phi_1-\psi_2^*\phi_2).
  \]
  The first of these quantities reduces to the fermion number of the standard (Grassmannian)
  Thirring model if we set $\phi=\psi$, while the second one vanishes.
\end{remark}

\section{Canonical formulation}\label{sec.CF}

In this section we shall present the canonical (Hamiltonian) formulation of the coupled massive
Thirring model defined by the field equations~\eqref{eqsm}. To begin with, the Lagrangian
generating these equations is given by Eq.~\eqref{L} with $m=-g=1$, or more explicitly
\[
  \cL=\frac\iu2\Big[\phi^*_a\dot\psi_a+\psi^*_a\dot\phi_a
        -(-1)^{a}\Big(\phi^*_a\psi_{a,x}+\psi_a^*\phi_{a,x}\Big)\Big]
            -\phi^*_1\psi_2-\phi^*_2\psi_1-2\phi^*_1\psi_1\phi^*_2\psi_2
         +\text{c.c.},
\]
where the dot denotes partial derivative with respect to the time $t$, c.c.~stands for the complex
conjugate, and (as in what follows) summation over repeated indices $a=1,2$ is understood. In
fact, in order to derive the canonical formulation of the field equations it is more convenient to
work with the equivalent (complex) Lagrangian
\begin{equation}
  \label{noncovL} 
  \cL=\iu\Big[\phi^*_a\dot\psi_a+\psi^*_a\dot\phi_a
  -(-1)^{a}\Big(\phi^*_a\psi_{a,x}+\psi_a^*\phi_{a,x}\Big)\Big]%\\
  -\Big(\phi^*_1\psi_2+\phi^*_2\psi_1+2\phi^*_1\psi_1\phi^*_2\psi_2
  +\text{c.c}\Big),
\end{equation}
which differs from the previous one by a trivial spacetime divergence. We shall next construct the
model's Hamiltonian following the canonical formalism for Grassmann-valued field theories outlined
in Refs.~\cite{Ca76,Ma85}. We start by defining the canonical momenta associated to the field
variables $(\chi,\chi^*)$, with
\[
\chi:=(\phi,\psi)=(\chi_\al)_{1\le\al\le4},
\]
as
\[
  \pi_{\chi_\al}=\frac{\pd\cL}{\pd\dot\chi_\al},\qquad
  \pi_{\chi^*_\al}=\frac{\pd\cL}{\pd\dot\chi^*_\al},
\]
where the partial derivatives, as in the sequel, are \emph{left} derivatives. With this convention
the Hamiltonian is defined as
\begin{equation}\label{Hdef}
  \cH=\dot\chi_\al\pi_{\chi_\al}+\dot\chi^*_\al\pi_{\chi^*_\al}-\cL,
\end{equation}
where summation over repeated indices $\al=1,\dots,4$ is again understood. In our case we have
\begin{equation}
  \label{pichia}
  \pi_{\phi_a}=-\iu\psi_a^*,\qquad \pi_{\psi_a}=-\iu\phi_a^*,
  \qquad \pi_{\phi_a^*}=
  \pi_{\psi_a^*}=0,
\end{equation}
and thus the Hamiltonian of the model is simply
\begin{equation}
  \cH=\dot\chi_\al\pi_{\chi_\al}-\cL=\iu
  (-1)^{a}\Big(\phi^*_a\psi_{a,x}+\psi_a^*\phi_{a,x}\Big)
  +\Big(\phi^*_1\psi_2+\phi^*_2\psi_1+2\phi^*_1\psi_1\phi^*_2\psi_2
  +\text{c.c.}\Big).
  \label{cH}  
\end{equation}
At this point it is important to note that Eqs.~\eqref{pichia} cannot be used (as is usual with
bosonic field theories) to express the generalized velocities $\dot\chi_\al,\dot\chi^*_\al$ in
terms of the corresponding canonical momenta, but are rather \emph{constraints} relating the
canonical momenta to the field variables. Thus in this case the canonical formalism should be
developed using Dirac's method for constrained systems~\cite{Di50} as presented, e.g., in
Ref.~\cite{Da21}. However, the vanishing of the canonical momenta associated to the conjugate
field variables $\chi_\al^*$ suggests that we can simply regard the fields $\chi_\al$ and their
conjugate momenta $\pi_{\chi_\al}$ as the fundamental canonical variables, using the first two
equations in~\eqref{pichia} to relate the complex conjugate fields~$\chi^*_\al$ to the canonical
momenta $\pi_{\chi_\al}$ (see the appendix for a detailed justification of this statement). Taking
into account our convention of using left derivatives, it is then easily checked that the field
equations~\eqref{eqsm} adopt indeed the canonical form
\begin{equation}\label{FE}
  \dot\chi_\al=-\frac{\de\cH}{\de\pi_{\chi_\al}},\qquad
  \dot\pi_{\chi_\al}=-\frac{\de\cH}{\de\chi_\al},
\end{equation}
where $(\frac{\de\cH}{\de\chi_\al},\frac{\de\cH}{\de\pi_{\chi_\al}})$ are variational derivatives
defined by the relation\footnote{In this section we shall often drop the time dependence of the
  fields and the canonical momenta, writing for instance $\chi_\al(x)$ instead of
  $\chi_\al(x,t)$.}
\[
  \de H=\int\diff x\bigg(\de\chi_\al\,\frac{\de\cH}{\de\chi_\al(x)} +\de\pi_{\chi_\al}\frac{\de
    \cH}{\de\pi_{\chi_\al}(x)}\bigg),\qquad \text{with}\quad H:=\int\diff x\,\cH.
\]

Following Refs.~\cite{Ca76,Ma85}, we define the Poisson bracket of two dynamical variables
$F[\chi_\al,\pi_{\chi_\al}]=\int\diff x\,\cF$ and $G[\chi_\al,\pi_{\chi_\al}]=\int\diff x\,\cG$ by
\begin{equation}
  \label{PB}
  \big\{F,G\}=(-1)^{|F|}\int\diff x\,\bigg(\frac{\de\cF}{\de\chi_\al(x)}\frac{\de
    \cG}{\de\pi_{\chi_\al}(x)}
  +\frac{\de\cF}{\de\pi_{\chi_\al}(x)}\frac{\de
    \cG}{\de\chi_\al(x)}
  \bigg)=:\int\diff x\,\{\cF,\cG\},
\end{equation}
where $|F|$ is the grading of $F$ (i.e., $|F|=0$ if $F$ is even and $|F|=1$ if $F$ is odd). With
this definition the fundamental Poisson brackets are given by\footnote{Recall that the Poisson
  bracket between two odd dynamical functions is symmetric.}
\begin{equation}
    \label{FPB}
\begin{aligned}
  &\{\chi_\al(x),\chi_\be(y)\}=\{\pi_{\chi_\al}(x),\pi_{\chi_\be}(y)\}=0,\\
  &\{\chi_\al(x),\pi_{\chi_\be}(y)\}=\{\pi_{\chi_\al}(x),\chi_\be(y)\}=-\de_{\al\be}\de(x-y).
\end{aligned}
\end{equation}
(Note the minus sign, which is due to our choice of left derivatives.)
Using Eqs.~\eqref{pichia} we obtain the fundamental Poisson brackets
\begin{equation}\label{FPBexp}
  \{\phi_{a}(x),\psi^*_b(y)\}=\{\psi^*_b(y),\phi_{a}(x)\}=\{\psi_{a}(x),\phi^*_b(y)\}
  =\{\phi^*_b(y),\psi_{a}(x)\}=-\iu\,\de_{ab}\de(x-y).
\end{equation}
In particular, the local first integrals computed in the previous sections are of the form
\begin{equation}\label{Ftype}
  F=\int\diff x\,\cF(\chi_\al,\chi_\al^*,\pd_x\chi_\al,\pd_x\chi^*_\al,\dots,\pd_x^n\chi_\al,
  \pd_x^n\chi^*_\al),
\end{equation}
where as remarked above the complex conjugates of the fields are related to the canonical momenta
by the first two equations in~\eqref{pichia}. Taking this into account, the Poisson bracket
between two (even) dynamical variables $F_1$ and $F_2$ of the latter type is explicitly given by
Eq.~\eqref{PB} with
\begin{equation}\label{PBF}
  \big\{\cF_1,\cF_2\big\}=\iu\bigg(\frac{\de \cF_1}{\de\phi_a(x)}%
  \frac{\de \cF_2}{\de\psi^*_a(x)}+\frac{\de \cF_1}{\de\psi_a(x)}%
  \frac{\de \cF_2}{\de\phi^*_a(x)}
  +\frac{\de \cF_1}{\de\phi^*_a(x)}%
  \frac{\de \cF_2}{\de\psi_a(x)}+\frac{\de \cF_1}{\de\psi^*_a(x)}%
  \frac{\de \cF_2}{\de\phi_a(x)}
  \bigg),
\end{equation}
where
\begin{equation}\label{vdFi}
  \frac{\de \cF_i}{\de\chi_a}:=\frac{\pd\cF_i}{\pd\chi_a}
  +\sum_{k=1}^n(-1)^k\pd_x^k\frac{\pd\cF_i}{\pd(\pd_x^k\chi_a)},\qquad i=1,2
\end{equation}
and similarly for $\frac{\de\cF_i}{\de\chi^*_a}$.

\smallskip The action $\int\diff t\,\diff x\,\cL$ of the Lagrangian~\eqref{noncovL} is manifestly
invariant under constant spacetime translations
\begin{equation}\label{trans}
  x^\mu\mapsto x'^\mu=x^\mu+\vep^\mu,\qquad\vep^\mu\in\RR,\quad\mu=0,1,
\end{equation}
as well as under global $\mathrm U(1)$ gauge transformations
\begin{equation}\label{gauge}
  \chi\mapsto\e^{\iu\vep}\chi,\qquad \chi^*\mapsto\e^{-\iu\vep}\chi^*,\qquad \vep\in\RR,
\end{equation}
and global $\SU(1,1)$ scaling transformation~\eqref{SU11symm}. Moreover, in view of Eq.~\eqref{L}
the action is also invariant under Lorentz boosts
\begin{equation}\label{Lorentz}
  (x,t)\mapsto(x',t')=(x\cosh\vep-t\sinh\vep,-x\sinh\vep+t\cosh\vep),\qquad\vep\in\RR.
\end{equation}
Indeed, in light-cone coordinates $\xi=(t+x)/2$, $\eta=(t-x)/2$ the Lorentz boost~\eqref{Lorentz}
becomes
\[
  (\xi,\eta)\mapsto(\xi',\eta')=(\e^{-\vep}\xi,\e^\vep\eta)
\]
and the Lagrangian~\eqref{noncovL} reads
\begin{align*}
  \cL&=\iu\Big(\phi^*_1\psi_{1,\xi}+\phi^*_2\psi_{2,\eta}
  +\psi^*_1\phi_{1,\xi}+\psi^*_2\phi_{2,\eta}\Big)
  -\Big(\phi^*_1\psi_2+\phi^*_2\psi_1+2\phi^*_1\psi_1\phi^*_2\psi_2
  +\text{c.c.}\Big)\\
      &=\iu\e^{-\vep}\Big(\phi^*_1\psi_{1,\xi'}+\psi^*_1\phi_{1,\xi'}\Big)
        +\iu\e^{\vep}\Big(\phi^*_2\psi_{2,\eta'}+\psi^*_2\phi_{2,\eta'}\Big)
       -\Big(\phi^*_1\psi_2+\phi^*_2\psi_1+2\phi^*_1\psi_1\phi^*_2\psi_2+\text{c.c.}\Big). 
\end{align*}
Hence the action will be invariant under the Lorentz boost~\eqref{Lorentz} provided that the
fields transform as
\begin{equation}
  \label{trfL}
  \phi(x,t)\mapsto\phi'(x',t')=\e^{-\frac\vep2\si_z}
  \phi(x,t),\qquad
  \psi(x,t)\mapsto\psi'(x',t')=\e^{-\frac\vep2\si_z}
  \psi(x,t),
\end{equation}
and similarly for $(\phi^*,\psi^*)$. Note, in particular, that each field component transforms
under a \emph{different} irreducible representation $(\e^{\pm\vep})$ of the Lorentz
group\footnote{Note that, since the Lorentz group in two spacetime dimensions is one-dimensional,
  all its irreducible representations are necessarily one-dimensional.}.

The conserved current associated to the invariance under spacetime translations by Noether's
theorem is the energy-momentum tensor
\[
  T^\mu{}_\nu=\chi_{\al,\nu}\frac{\pd\cL}{\pd\chi_{\al,\mu}}-\de^{\mu}{}_\nu\cL,\qquad
  \mu,\nu=0,1,
\]
where we have used the fact that the Lagrangian~\eqref{noncovL} does not contain derivatives of
the complex conjugates of the fields. The corresponding conserved quantities are
\[
  \int\diff x\,T^0{}_0=H,\qquad
  \int\diff x\,T^0{}_1=\iu\int\diff x
  \Big(\psi_a^*\phi_{a,x}+\phi^*_a\psi_{a,x}
  \Big)=:-P,
\]
where $P$ is the total linear momentum of the fields. Note that $H$ and $P$ can be easily
expressed in terms of the lower-order local conserved quantities computed in the previous section
as follows:
\begin{align*}
  H=-\frac12\Re\Big(I_3^{(1)}+\wt I_3^{(1)}\Big),\qquad P=\frac12\Re\Big(I_3^{(1)}-\wt I_3^{(1)}\Big).
\end{align*}
Likewise, from the global $\mathrm U(1)$ gauge invariance of the Lagrangian we deduce the
conservation of the current
\[
  j^\mu=-\iu\chi_\al\frac{\pd\cL}{\pd\chi_{\al,\mu}},\qquad\mu=0,1.
\]
The corresponding conserved quantity is the system's total charge (or fermion number)
\[
  Q_R:=\int\diff x\,j^0=\int\diff x\,\big( \phi^*_a\psi_a+\psi^*_a\phi_a\big)=2\Re\int\diff
  x\,\rho_+.
\]
Similarly, the invariance of the Lagrangian under the $\SU(1,1)$ transformation~\eqref{SU11symm}
yields the conserved current
\[
  j^\mu=\phi_a\frac{\pd\cL}{\pd\phi_{\al,\mu}}-\psi_a\frac{\pd\cL}{\pd\psi_{\al,\mu}},\qquad\mu=0,1,
\]
whose conserved quantity is given by
\[
  Q_I:=\int\diff x\,\iu\big(\phi_a^*\psi_a-\psi_a^*\phi_a\big)=-2\Im\int\diff x\,\rho_+.
\]
Finally, from the invariance of the action under the Lorentz boosts~\eqref{Lorentz}-\eqref{trfL}
it follows that the current
\[
  j^\mu= xT^\mu{}_0+tT^\mu{}_1+\frac{(-1)^a}2\bigg(\phi_a\frac{\pd\cL}{\pd\phi_{a,\mu}}
  +\psi_a\frac{\pd\cL}{\pd\psi_{a,\mu}}\bigg),
\]
is conserved (see, e.g., Ref.~\cite{Ma05}). The corresponding conserved quantity is given by
\begin{equation}\label{consLor}
  \int\diff x\,j^0=% \int\diff x\bigg[x T^0{_0}+t T^0{}_1+\frac\iu2(-1)^a\big(\phi_a^*\psi_a
  % +\psi_a^*\phi_a\big)\bigg]
  \int\diff x\bigg[xH+\frac\iu2(-1)^a\big(\phi_a^*\psi_a+\psi_a^*\phi_a\big)\bigg]-tP.
\end{equation}
Note that the time-dependent conservation law~\eqref{consLor} amounts to the equation of motion
\[
  \int\diff x\bigg[xH+\frac\iu2(-1)^a\big(\phi_a^*\psi_a+\psi_a^*\phi_a\big)\bigg]
  =tP+\text{const.}
\]
In particular, the left-hand side of the latter relation is conserved in the system's center of
momentum frame, in which $P=0$.
\begin{remark}
  Apart from the continuous symmetries discussed above, the Lagrangian~\eqref{L} is manifestly
  invariant under the following transformations:
  
  \setbox0=\hbox{ii)\en Time reversal $\cT$:}
  \count1=\wd0
  
  \medskip
  \begin{tabbing}
    \noi ii)\en Time reversal $\cT$:\quad\=\kill
    \noi \hphantom{i}i)\en Parity~$\cP$:\> $\ds (x,t)\to(-x,t),\quad
    \phi(x,t)\rightarrow\si_x\phi(-x,t)$,
    \quad $\psi(x,t)\rightarrow\si_x\psi(-x,t)$.\\[1mm]
    \noi ii)\en Time reversal $\cT$:\quad
    $\ds (x,t)\to(x,-t),\quad \phi(x,t)\rightarrow\si_x\phi^*(x,-t),
    \quad \psi(x,t)\rightarrow-\si_x\psi^*(x,-t)$.
  \end{tabbing}
  
  \medskip
  \noi In particular, the field equations~\eqref{eqsm} are invariant under the $\cP\cT$ mapping
  \[
    (x,t)\to (-x,-t),\quad \big(\phi(x,t),\psi(x,t)\big)
    \rightarrow\big(\phi^*(-x,-t),-\psi^*(-x,-t)\big).
  \]
\end{remark}

\medskip Using Eqs.~\eqref{PBF}-\eqref{vdFi}, it is straightforward to compute the Poisson
brackets of the lowest-order local conserved quantities $\int\diff x\,\rho_+=I_2^{(0)}$,
$I_3^{(1)}$, and $\wt I_3^{(1)}$ derived in the previous section. To begin with, if $F$ is of the
form~\eqref{Ftype} we have
\[
  \Big\{I_2^{(0)},F\Big\}=\iu\int\diff x\,\bigg(\psi_a(x)\frac{\de
    \cF}{\de\psi_a(x)}-\phi^*_a(x)\frac{\de \cF}{\de\phi^*_a(x)}\bigg).
\]
For $F=I_3^{(1)}$, the integrand in the RHS of this equation is (omitting the argument of the
fields)
\begin{multline*}
  -2\Big[\psi_1(-\phi^*_2-2\phi^*_1\phi^*_2\psi_2)+\psi_2(2\iu\phi^*_{2,x}
  -\phi_1^*-2\phi^*_1\phi_1\phi_2^*)\\-\phi^*_1(\psi_2+2\psi_1\phi^*_2\psi_2)
  -\phi^*_2(2\iu\psi_{2,x}+\psi_1+2\phi^*_1\psi_1\psi_2)
  \Big]=4\iu\Big(\phi^*_2\psi_{2,x}-\psi_2\phi^*_{2,x}\Big)=4\iu\,\pd_x\rho_2,
\end{multline*}
which vanishes upon integration. Thus
\[
  \Big\{I_2^{(0)},I_3^{(1)}\Big\}=0,
\]
and a similar calculation shows that $\{I_2^{(0)},\wt I_3^{(1)}\}$ vanishes as well. Finally, if
we denote by $\cI_3^{(1)}$ and $\wt\cI_3^{(1)}$ the densities of the first integrals $I_3^{(1)}$
and $\wt I_3^{(1)}$ from Eqs.~\eqref{I3112}-\eqref{tI31} it easily follows that
\begin{align*}
  \Big\{\cI_3^{(1)},\wt \cI_3^{(1)}\Big\}
  &=8\iu\Big\{\phi_a^*\psi_{a,x},\phi^*_1\psi_2+\phi^*_2\psi_1
    +2\phi^*_1\psi_1\phi^*_2\psi_2\Big\}\\
  &=-8\bigg[\psi_{1,x}(-\phi^*_2-2\phi^*_1\phi^*_2\psi_2)
    +\psi_{2,x}(-\phi^*_1-2\phi^*_1\psi_1\phi^*_2)+\phi^*_{1,x}(\psi_2+2\psi_1\phi^*_2\psi_2)\\
  &+\phi^*_{1,x}(\psi_2+2\psi_1\phi^*_2\psi_2)
    +\phi^*_{2,x}(\psi_1+2\phi^*_1\psi_1\psi_2)\bigg]
    =-8\pd_x\Big(\phi^*_1\psi_2+\phi^*_2\psi_1+2\rho_1\rho_2\Big),
\end{align*}
and hence
\[
  \Big\{I_3^{(1)},\wt I_3^{(1)}\Big\}=0.
\]
The previous explicit calculations show that the conserved quantities $I_2^{(0)}$, $I_3^{(1)}$,
$\wt I_3^{(1)}$ derived in the previous section Poisson commute among each other. In fact, with
the help of \emph{Mathematica}\texttrademark{} we have checked that $I_3^{(3)}$ and
$\wt I_3^{(3)}$ are also in involution with the latter quantities and among themselves. This
strongly suggests that the coupled Thirring model~\eqref{eqsm} is completely integrable in
Liouville's sense, as was shown to be the case for its bosonic counterpart~\cite{BS23}. The
analysis of this conjecture will in fact be the subject of a forthcoming publication.

\section{Nonlocal reductions and their symmetries}\label{sec.NRS}

In the previous sections we have regarded the fields $\phi$ and $\psi$ as independent, obeying the
coupled field equations~\eqref{eqsm}. However, as is the case with the bosonic version of the
model studied in Ref.~\cite{BS23}, these equations admit several interesting reductions that we
shall now analyze in detail.

As mentioned in Section~\ref{sec.Lax}, the simplest reduction of Eqs.~\eqref{eqsm} is obtained
setting $\phi=\psi$, in which case the $\psi$ (or $\phi$ field) obeys the equations of the
original (Grassmannian) Thirring model. More precisely, when $\phi=\psi$ the Lagrangian~\eqref{L}
reduces to (twice) the Lagrangian of the ordinary Grassmannian MTM, and similarly for the
corresponding Hamiltonian. Likewise, the Lax pair and the integrals of motion of the coupled GMTM
turn into the Lax pair and the integrals of motion of the ordinary GMTM under the replacement of
$\phi$ by $\psi$. Apart from this trivial reduction, only two of the five nonlocal reductions of
the coupled bosonic MTM studied in Ref.~\cite{BS23} survive in our case, namely the real
space/time reflections defined by

\begin{tabbing}
  \noi\hphantom{I}I)\en Real space reflection:\quad\=
  $\ds x\mapsto -x$,\quad\=$\phi(x,t)= \si_x\psi(-x,t)$\\[3pt]
  \noi II)\en Real time reflection:\>$\ds t\mapsto -t$,
  \>$\phi(x,t)= -\iu\si_y\psi(x,-t)$
\end{tabbing}

\smallskip\noi The equations of motion of the $\psi$ field in each of these reductions are
respectively
\begin{equation}\label{typeI}
\begin{aligned}
  &\iu(\pd_t+\pd_x)\psi_1-\psi_2+2\psi^*_1(-x,t)\psi_1\psi_2=0,\\
  &\iu(\pd_t-\pd_x)\psi_2-\psi_1+2\psi^*_2(-x,t)\psi_2\psi_1=0
\end{aligned}
\end{equation}
and
\begin{equation}\label{typeII}
\begin{aligned}
  &\iu(\pd_t+\pd_x)\psi_1-\psi_2+2\psi^*_1(x,-t)\psi_1\psi_2=0,\\
  &\iu(\pd_t-\pd_x)\psi_2-\psi_1-2\psi^*_2(x,-t)\psi_2\psi_1=0,\\
\end{aligned}
\end{equation}
where for the sake of simplicity we have suppressed the usual argument $(x,t)$ wherever
appropriate. In both cases, the Lax pair and the integrals of motion are obtained from those of
the coupled GMTM replacing the $\phi$ field by its expression in terms of $\psi(-x,t)$ or
$\psi(x,-t)$, and the same is true for the Lagrangian and the corresponding Hamiltonian. For
instance, the Lax pair for Eqs.~\eqref{typeI}-\eqref{typeII} is given by Eq.~\eqref{u}-\eqref{uvn}
with $\rho_\pm$ and $r_1^\pm$ defined by
\begin{tabbing}
  \noi\hphantom{I}I)\quad\=
  $\ds \rho_\pm=\psi_1^*(-x,t)\psi_2\pm\psi_2^*(-x,t)\psi_1,\qquad
  r^\pm_1=-\iu\big[\la\psi_1^*(-x,t)\pm\la^{-1}\psi_2^*(-x,t)\big],$\\[3pt]
  \noi II)\>$\rho_\pm=\psi_1^*(-x,t)\psi_2\mp\psi_2^*(-x,t)\psi_1,\qquad
  r^\pm_1=-\iu\big[\la\psi_1^*(-x,t)\mp\la^{-1}\psi_2^*(-x,t)\big]$.
\end{tabbing}

The above reductions clearly remain invariant under spacetime translations~\eqref{trans} and
global $U(1)$ gauge transformations~\eqref{gauge}. It should be noted, however, that these
reductions break the Lorentz invariance of the original coupled model~\eqref{L}. This is
ultimately due to the fact that in both of them the second (resp.~first) component of the $\phi$
field is related to the first (resp.~second) component of the $\psi$ field, whereas according to
Eq.~\eqref{trfL} these components transform under different irreducible representations of the
Lorentz group. The breakdown of Lorentz invariance can also be directly checked from the field
equations, which in light-cone coordinates can be written in terms of the transformed variables
$(\xi',\eta')=(\e^{-\vep}\xi,\e^{\vep}\eta)$ as
\begin{tabbing}
  \noi\hphantom{I}I)\quad\=
  $\iu\e^{-\vep}\psi_{1,\xi'}-\psi_2+2\psi_1^*(\eta,\xi)\psi_1\psi_2=0,\qquad
  \iu\e^{\vep}\psi_{2,\eta'}-\psi_1+2\psi_2^*(\eta,\xi)\psi_2\psi_1=0,$\\[3pt]
  \noi II)\>$\iu\e^{-\vep}\psi_{1,\xi'}-\psi_2+2\psi_1^*(-\eta,-\xi)\psi_1\psi_2=0,\qquad
  \iu\e^{\vep}\psi_{2,\eta'}-\psi_1-2\psi_2^*(-\eta,-\xi)\psi_2\psi_1=0.$
\end{tabbing}
It is clear that no scaling transformation $(\psi_1,\psi_2)=(\la_1\psi_1',\la_2\psi_2')$ can
eliminate the $\e^{\pm\vep}$ factors in both equations of motion.

On the other hand, it is straightforward to show that both nonlocal
reductions~\eqref{typeI}-\eqref{typeII} preserve the invariance under space/time reflections of
the original equations~\eqref{eqsm}. Indeed, the type I) field equations~\eqref{typeI} are easily
seen to be invariant under the transformations
\begin{equation}\label{PT-I}
\begin{alignedat}{4}
  \cP&:\qquad x&&\to-x,\qquad &&\psi(x,t)&&\to\si_x\psi(-x,t),\\
  \cT&:\qquad t&&\to-t,&&\psi(x,t)&&\to-\iu\si_y\psi(x,-t).
\end{alignedat}
\end{equation}
Likewise, the type II) reduction equations of motion~\eqref{typeII} are invariant under
\begin{equation}\label{PT-II}
\begin{alignedat}{4}
  \cP&:\qquad x&&\to-x,\qquad &&\psi(x,t)&&\to\iu\si_y\psi^*(-x,t),\\
  \cT&:\qquad t&&\to-t,&&\psi(x,t)&&\to\si_x\psi^*(x,-t).
\end{alignedat}
\end{equation}
In particular, both reductions~\eqref{typeI}-\eqref{typeII} are $\cP\cT$-symmetric, i.e.,
invariant under the composition of the mappings $\cP$ and $\cT$, which in both cases is explicitly
given by
\begin{equation}\label{cPT}
\cP\cT:\qquad (x,t)\to(-x,-t),\qquad\psi(x,t)\to\si_z\psi(-x,-t).
\end{equation}

\section{Conclusions and outlook}\label{sec.CO}

In this paper we introduce a model of two interacting Dirac fermions with Thirring
self-interactions whose fields take values in a Grassmann algebra, which can be regarded as the
fermionic version of the bosonic model recently studied in Ref.~\cite{BS23}. The model is
relativistically invariant, and is in addition symmetric under space and time reflections and
$\mathrm U(1)$ global gauge transformations. It contains as particular cases the free Dirac
equation (when its coupling constant vanishes) or the Grassmannian massive Thirring model (when
its two independent field variables are set to be equal). We start by constructing a Lax pair for
the model, from which the field equations are derived as a zero curvature condition. We also show
that the model is closely related to an $\SU(1,1)$ version of the Grassmannian Thirring model that
we introduce in this work. Following the approach of Ref.~\cite{Ag12} for the Grassmannian
Thirring model, from the Riccati-type equations satisfied by the quotient of suitable components
of the auxiliary vector variable in the Lax pair we obtain four infinite hierarchies of conserved
quantities. All of these quantities turn out to be nonlocal, with the exception of the two
lowest-order ones. A variant of this method, going back to Refs.~\cite{IK83,IK78}, is then used to
construct four additional infinite hierarchies of local conserved quantities.

Using Dirac's method for constrained systems, we develop in detail the Hamiltonian formulation of
the model. As is the case with the Dirac equation, the definitions of the canonical momenta turn
out to be second-class constraints allowing the determination of all the Lagrange multipliers
associated to the constraints. A consistent Poisson bracket for the system can then be constructed
by eliminating the complex conjugates of the fields and the corresponding canonical momenta. Using
this bracket, we show that the first few lower-order local first integrals previously computed
Poisson commute with each other, which strongly suggests that the model is completely integrable
in Liouville's sense.

Following the work in Ref.~\cite{AM13} on nonlocal integrable reductions of the nonlinear
Schrödinger equation, and the analogous results for the bosonic coupled Thirring model in
Ref.~\cite{BS23}, we investigate the existence of such reductions for the present model. We
construct two nonlocal integrable reductions based on real space and time reflections, and
investigate their symmetries. Although the Lorentz invariance of the general model is not
preserved, we show that both of these reductions are invariant under parity ($\cP$) and time
reversal ($\cT$), and are thus $\cP\cT$-invariant. We also show that the Lax pair, Lagrangian,
Hamiltonian and hierarchies of conserved quantities of the nonlocal reductions can be easily
constructed by suitable reductions of their counterparts for the general model.

The present work suggests several lines for future work. In the first place, it should be of
interest to study the complete integrability of the model by a suitable generalization of the
method used in Ref.~\cite{BS23} to establish this property for its bosonic counterpart. Another
natural problem is to investigate the existence of solitonic solutions ---which the standard
Grassmannian Thirring model does not possess~\cite{IK78}--- using the inverse scattering method.
Likewise, it would be of interest to construct particular solutions of the model using Bäcklund
transformations, as done in Ref.~\cite{IS76} for the Grassmannian Thirring model. A natural
continuation of the present work would be the construction of the quantized version of the model
and the analysis of its properties (integrability, spectrum, symmetries, etc), including its
unitarity. Finally, it would also be of interest to investigate the integrability properties of
the $\SU(1,1)$ generalization of the GMTM introduced in this paper. Work on some of this topics is
currently going on and will appear in future publications.

\appendix

\section{Derivation of the fundamental Poisson brackets~\eqref{FPB} through Dirac's method
  for constrained systems}\label{app.Dirac}

In this appendix we shall apply Dirac's method for systems with constraints to justify
Eq.~\eqref{FPB} for the fundamental Poisson brackets of the coupled Thirring model used in
Section~\ref{sec.CF}. Our starting point is the Lagrangian~\eqref{noncovL}, which leads to the
expressions~\eqref{pichia} for the conjugate momenta $(\pi_{\chi},\pi_{\chi^*})$ of the field
variables $(\chi,\chi^*)\equiv(\phi,\psi,\phi^*,\psi^*)$. Since these expressions do not involve
the generalized velocities $(\dot\chi,\dot\chi^*)$, they give rise to the set of primary
constraints
\begin{equation}
  \label{const}
  \Ga\equiv(\Ga_A)_{1\le A\le 8}=(\pi_{\phi}+\iu\psi^*,\pi_{\psi}+\iu\phi^*,
  \pi_{\phi^*},\pi_{\psi^*}).
\end{equation}
Following Dirac's method, we then define the primary Hamiltonian density
\[
  \cH_P=\cH+\la_A\Ga_A,
\]
where $\cH$ is the canonical Hamiltonian~\eqref{cH} obtained from Eq.~\eqref{Hdef}, and the
$\la_A$ are (odd) Lagrange multipliers depending on the spacetime coordinates $(x,t)$ but
independent of the field variables and their conjugate momenta. The canonical equations of motion
generated by the primary Hamiltonian
\[
  H_P=\int\diff x\,\Big(\cH(x)+\la_A(x)\Ga_A(x)\Big)=H+\int\diff x\,\la_A(x)\Ga_A(x)
\]
are the most general equations of motion compatible with variations of the fields and their
conjugate momenta respecting the constraints. As is the case for the Dirac equation (see, e.g.,
Ref.~\cite{Da21}), all the primary constraints turn out to be second class. In other words,
imposing that the primary constraints be consistent with the time evolution, i.e.,
that\footnote{As is customary, we shall use the symbol $\approx$ to denote an equality modulo the
  constraints.}
\begin{equation}\label{canPB}
  \{\Ga_A,H_P\}\approx0,\qquad 1\le A\le 8,
\end{equation}
determines the Lagrange multipliers $\la_A$. It should be noted that the Poisson bracket in
Eq.~\eqref{canPB} is the \emph{canonical} one, computed regarding the fields $(\chi,\chi^*)$ and
their conjugate momenta $(\pi_\chi,\pi_{\chi^*})$ as \emph{independent} (anticommuting) variables
satisfying the standard canonical relations\footnote{Throughout this appendix we shall omit the
  time variable, whose value is the same for all fields and momenta.}
\[
  \{\chi_\al(x),\pi_{\chi_\be}(y)\}=\{\pi_{\chi_\al}(x),\chi_\be(y)\}
  =\{\chi_\al^*(x),\pi_{\chi_\be^*}(y)\}=\{\pi_{\chi_\al^*}(x),\chi_\be^*(y)\}=-\de_{\al\be}\de(x-y)
\]
(all other Poisson brackets vanishing identically). For instance, from the Poisson bracket
\begin{align*}
  \{\Ga_1(x),H_P\}
  &=\bigg\{\pi_{\phi_1}(x)+\iu\psi_1^*(x),H+\int\diff y\,\la_A(y)\Ga_A(y)\bigg\}\\
  &=\{\pi_{\phi_1}(x),H\}-\int\diff y\,\la_7(y)\Big\{\pi_{\phi_1}(x)+\iu\psi_1^*(x),
    \pi_{\psi_1^*}(y)\Big\}\\
  &=-\iu\psi_{1,x}^*(x)-\psi_2^*(x)-2\psi_1^*(x)\psi_2^*(x)\phi_2(x)+\iu\la_7(x)
\end{align*}
we obtain
\[
  \la_7=\psi_{1,x}^*-\iu\Big(\psi_2^*+2\psi_1^*\psi_2^*\phi_2\Big).
\]
Proceeding in this way we arrive at the following expressions for the Lagrange multipliers
$\la_A(x)$:
\begin{alignat*}{2}
  \la_1&=-\phi_{1,x}-\iu\Big(\phi_2+2\phi_1\psi_2^*\phi_2\Big),\qquad
  &\la_2&=\phi_{2,x}-\iu\Big(\phi_1+2\phi_2\psi_1^*\phi_1\Big),\\
\la_3&=-\psi_{1,x}-\iu\Big(\psi_2+2\psi_1\phi_2^*\psi_2\Big),\qquad
  &\la_4&=\psi_{2,x}-\iu\Big(\psi_1+2\psi_2\phi_1^*\psi_1\Big),\\
       \la_5&=\phi_{1,x}^*-\iu\Big(\phi_2^*+2\phi_1^*\phi_2^*\psi_2\Big),
  &\la_6&=-\phi_{2,x}^*-\iu\Big(\phi_1^*+2\phi_2^*\phi_1^*\psi_1\Big),\\
  \la_7&=\psi_{1,x}^*-\iu\Big(\psi_2^*+2\psi_1^*\psi_2^*\phi_2\Big),
  &\la_8&=-\psi_{2,x}^*-\iu\Big(\psi_1^*+2\psi_2^*\psi_1^*\phi_1\Big).
\end{alignat*}
As expected, the Lagrange multipliers coincide with the generalized velocities
$(\dot\chi,\dot\chi^*)$ expressed in terms of the fields and their space derivatives
(cf.~Eqs.~\eqref{eqsm}). Furthermore, since Eqs.~\eqref{canPB} have not produced any new
constraints the primary constraints~\eqref{const} are the only constraints in the system. In
addition, since all the Lagrange multipliers have been determined the Dirac matrix with elements
\[
  C_{AB}(x,y):=\{\Ga_A(x),\Ga_B(y)\},\qquad 1\le A,B\le 8,
\]
is invertible, and the Dirac bracket of the densities $\cF(x),\cG(y)$ of any two dynamical
variables is given by the standard formula
\begin{equation}
  \label{DB}
  \{\cF(x),\cG(y)\}_D=\{\cF(x),\cG(y)\}-\int\diff z\diff w\,\{\cF(x),\Ga_A(z)\}(C^{-1})_{AB}(z,w)
  \{\Ga_B(w),\cG(y)\}.
\end{equation}
In our case the Dirac matrix is easily computed, with the result\footnote{Note that in this case
  the Dirac matrix is not antisymmetric but rather \emph{symmetric}, as the Poisson bracket of two
  odd variables is symmetric.}
\[
  C(x,y)=-\iu\de(x,y)
  \begin{pmatrix}
    0&0&0&\id\\
    0&0&\id&0\\
    0&\id&0&0\\
    \id&0&0&0
  \end{pmatrix},
\]
where $\id$ denotes the $2\times2$ identity matrix. For instance,
\[
  \{\Ga_1(x),\Ga_B(y)\}=\{\pi_{\phi_1}(x)+\iu\psi_1^*(x),\Ga_B(y)\}
  =\de_{B7}\{\iu\psi_1^*(x),\pi_{\psi_1^*}(y)\}=-\iu\de_{B7}\de(x-y).
\]
Since $C^{-1}(x,y)=-C(x,y)$, from Eq.~\eqref{DB} we immediately obtain the following explicit
expression for the Dirac bracket:
\begin{align}
  \{\cF(x),\cG(y)\}_D=\{\cF(x),\cG(y)\}-\iu\int\diff z
  &\bigg[\{\cF(x),\pi_{\phi_a}(z)+\iu\psi_a^*(z)\}
    \{\pi_{\psi_a^*}(z),\cG(y)\}\notag\\
  &+\{\cF(x),\pi_{\psi_a}(z)+\iu\phi_a^*(z)\}
    \{\pi_{\phi_a^*}(z),\cG(y)\}\vphantom{\bigg]}\notag\\
  &+\{\cF(x),\pi_{\phi_a^*}(z)\}
    \{\pi_{\psi_a}(z)+\iu\phi_a^*(z),\cG(y)\}\notag\\
  &+\{\cF(x),\pi_{\psi_a^*}(z)\}
    \{\pi_{\phi_a}(z)+\iu\psi_a^*(z),\cG(y)
    \bigg].
    \label{DBexp}
\end{align}
Using Eq.~\eqref{DBexp} it is straightforward to show that the only nonvanishing Dirac brackets
between the fields $(\chi,\chi^*)$ appearing in the canonical Hamiltonian~\eqref{cH} turn out to
be
\begin{equation}\label{FDB}
  \{\phi_a(x),\psi_b^*(y)\}_D=\{\psi_a(x),\phi_b^*(y)\}_D=\{\phi_a^*(x),\psi_b(y)\}_D
  =\{\psi_a^*(x),\phi_b(y)\}_D=-\iu\de_{ab}\de(x-y).
\end{equation}
Indeed,
\begin{align*}
  \{\phi_a(x),\psi_b^*(y)\}_D&=\{\phi_a(x),\psi_b^*(y)\}-\iu\int\diff
  z\{\phi_a(x),\pi_{\phi_c}(z)+\iu\psi_c^*(z)\}\{\pi_{\psi_c^*}(z),\psi_b^*(y)\}\\
  &=-\iu\int\diff z(-\de_{ac}\de(z-x))(-\de_{bc}\de(z-y))=-\iu\de_{ab}\de(x-y),
\end{align*}
etc. Comparing with Eqs.~\eqref{FPBexp} we conclude that the prescription used in
Section~\ref{sec.CF} of replacing $\chi_\al^*$ by its expression~\eqref{pichia} in terms of the
canonical momenta $\pi_{\chi_\al}$ in the computation of Poisson brackets of dynamical variables
of the form~\eqref{Ftype} ---which amounts to using the constraints to eliminate the canonical
variables $(\chi^*,\pi_{\chi^*})$--- is justified if we interpret the Poisson bracket as a Dirac
bracket. Note, in this respect, that by construction all the constraints $\Ga_A(x)$ have vanishing
Dirac bracket with any dynamical variable $\cF(x)$, i.e.,
\[
  \{\cF(x),\Ga_A(y)\}_D=0,
\]
and therefore
\[
  \{\cF(x),H_P\}_D=\{\cF(x),H\}_D.
\]
Thus we can use the canonical Hamiltonian $H$ instead of the primary one $H_P$ to compute the time
evolution of dynamical variables, as we did in Section~\ref{sec.CF}. In particular, the fields'
equations of motion can be written in the canonical form~\eqref{FE}.

\acknowledgments The authors would like to thank the anonymous referee for his/her helpful
suggestions. This work was partially supported by grant~G/6400100/3000 from Universidad
Complutense de Madrid. DS acknowledges a research fellowship from CSIR (ACK No.: 362103/2k19/1,
File No.~09/489(0125)/2020-EMR-I), India.

% \bibliographystyle{JHEP}
% \bibliography{cmprefs}

\providecommand{\href}[2]{#2}\begingroup\raggedright\endgroup

\end{document}